\documentclass [a4paper]{article}

\usepackage[hiresbb]{graphicx}

\usepackage{setspace}
\usepackage{float}
\usepackage[title]{appendix}
\usepackage{nameref,hyperref}
\usepackage[table]{xcolor}
\usepackage{seqsplit}
\usepackage{lscape}
\usepackage{longtable}
\usepackage{rotating}
\usepackage{caption}
\usepackage{subcaption}
\usepackage{amsmath}
\usepackage{bm}

\begin{document}
\fontsize{10.5pt}{18pt}\selectfont

\fontsize{13pt}{18pt}\selectfont
\begin{center}
{\bf Lockdowns need geographic coordination\\ because of propagation of economic effects through supply chains
}

\vspace{2ex}

\fontsize{11pt}{18pt}\selectfont
Hiroyasu Inoue\footnote{Graduate School of Simulation Studies, University of Hyogo, inoue@sim.u-hyogo.ac.jp.}
Yohsuke Murase\footnote{RIKEN Center for Computational Science, yohsuke.murase@gmail.com}
Yasuyuki Todo\footnote{Graduate School of Economics, Waseda University, yastodo@waseda.jp.}
\vspace{3ex}

\fontsize{10.5pt}{18pt}\selectfont
Abstract
\end{center}

\vspace{1ex}
\fontsize{10pt}{12pt}\selectfont

\noindent In order to prevent the spread of COVID-19, governments have often required regional or national lockdowns, which have caused extensive economic stagnation over broad areas as the shock of the lockdowns has diffused to other regions through supply chains. Using supply-chain data for 1.6 million firms in Japan, this study examines how governments can mitigate these economic losses when they are obliged to implement lockdowns. Through tests of all combinations of two-region lockdowns, we find that coordinated, i.e., simultaneous, lockdowns yield smaller GDP losses than uncoordinated lockdowns. Furthermore, we test practical scenarios in which Japan's 47 regions
impose lockdowns over three months and find that GDP losses are lower if nationwide lockdowns are coordinated than if they are uncoordinated.

\vspace{1ex}

\noindent {\it Keywords}: COVID-19; lockdown; coordination; supply chains; simulation; propagation; 




\newpage

\fontsize{10pt}{10pt}\selectfont

\section{Introduction}

In order to prevent the spread of COVID-19 (SARS-CoV-2) in 2020,
many governments have imposed ``lockdowns'' on cities, regions, or entire countries, shutting down or shrinking economic and social activities~\cite{Hale2020}.
As of January, 2021, while some countries and regions have lifted lockdown, others continue or have resumed lockdown because of the second and third waves of the epidemic.
These lockdowns obviously decline the economic activity of locked-down regions.
In addition, the negative economic effect of lockdown diffuses 
through supply chains, i.e., supplier-client relationships of firms, 
to other regions that are not necessarily locked down.
When a firm shrinks its production due to lockdown, 
its client firms suffer from production reduction
because of lack of supply of intermediate goods and services.
Its suppliers also reduce production because of demand shortages.

Propagation of economic shocks due to natural disasters through supply chains is evidenced in the literature \cite{Barrot2016, Boehm2019, Carvalho2016, Inoue2019, Inoue2019b, Kashiwagi2018}.
More recently, the economic effect of anti-COVID-19 lockdowns that incorporates the supply-chain propagation is estimated, using input--output (IO) linkages at the country-sector level~\cite{Bonadio2020, Guan2020, McCann2020, McKibbin2020} and supply chains at the firm level~\cite{Inoue2020}.

However, one important aspect not fully examined in the literature is interactions of lockdowns of different regions and countries and resulting needs for policy coordination, although inter-regional and international policy coordination is discussed in some other context~\cite{Kremer2007, Taylor2013}.
The importance of this aspect can be easily recognised from the experience of Sweden. 
Sweden, unlike other European countries, did not impose a strict lockdown in 2020 to minimise the economic effect. However, it experienced a 7.4\% reduction in gross domestic product (GDP) in the second quarter of 2020 compared with in the previous year. 
This reduction for Sweden is comparable to those in neighbouring countries that imposed lockdown (-6.2\% for Finland and -4.6\% for Norway, Eurostat), possibly because of propagation of negative effects from locked-down countries to Sweden through economic networks~\cite{NYT2020}.
This anecdote may indicate that, provided lockdowns in neighbour countries, the Swedish government should have followed them and imposed lockdown to minimise the human loss while maintaining the economic loss similar to that from the no-lockdown strategy. 

One important study
uses data for human mobility and confirmed COVID-19 cases in simulations
to show the necessity of coordination across countries in the context of nonpharmaceutical interventions, such as social-distancing policies and lockdown strategies~\cite{ruktanonchai2020assessing}.
The study finds that coordination, i.e., simultaneous implementation of interventions, is effective to greatly reduce infection.
However, no study has examined the need for lockdown strategies to alleviate its negative economic effect,
except for a study~\cite{Inoue2020b} that focuses on the effect of lifting, rather than imposing, lockdown.

To fill the gap,
this study investigates the effect of regional coordination of lockdowns.
For this purpose, 
we conduct a simulation analysis applying an agent-based model of production to actual supply chain data of 1.6 million firms in Japan. 
Agent-based models that incorporate the interactions of agents through networks have been widely used in social science recently~\cite{Axtell2006, LeBaron2008, Gomez2019}. 
Then, utilising parameter values calibrated from the actual experiences of past lockdowns in Japan in April and May, 2020,
we perform many virtual simulations addressing the economic effect of combinations of lockdowns in different regions.
In particular, we examine
(1) how a lockdown in one prefecture affects production in other prefectures,
(2) how the effect of a simultaneous lockdown in two prefectures differ from the effects of asynchronous lockdowns in the same prefectures, and
(3) how nationwide coordination of lockdowns affects production.

\section{Data}

This study uses data collected by Tokyo Shoko Research (TSR), particularly, the Company Information Database and Company Linkage Database. The former data set contains  attributes for each firm, including its address, industry classification, and sales, while the latter consists of clients and suppliers. 
We particularly use the data for 2016. The number of firms is 1,668,567 and the number of supply-chain links is 5,943,073. Our data cover most firms in Japan except for micro enterprises and most major supply-chain relationships between them. However, they lack information about relationships with firms located in foreign countries.

Because sales from each supplier to each client and final consumers is not available in the data, we estimate the sales using sectoral sales volumes taken from the 2015 Input-Output Tables for Japan~\cite{METI15}. In this estimation process, we classify firms into 187 industries according to the IO Tables and drop firms without sales information. As a result, the number of firms after the volume estimation is 966,627 whereas the number of links is 3,544,343. Supplementary Information \ref{ch:supch} provides the details of the data and the estimation process.

We can consider the supply chains as a network,
and the number of links that each firm has, or the degree,
follows a power-law distribution~\cite{Inoue2019},
as is commonly observed in many natural and societal networks~\cite{Barabasi16}.
The average number of steps between firms, 
or the average path length in terms of supply chain partners, is 4.8.
If we consider the number of firms in the network (966,627), 
it is surprisingly small.
Such a network is commonly called a small-world network~\cite{Watts98}.
In addition, with the same data set, previous studies~\cite{Inoue2019, Fujiwara10} find that 46--48\% of the firms are included in the giant strongly connected component (GSCC), in which all firms are directly or indirectly connected through directed links. In other words, approximately a half of firms are involved in numerous cycles in the GSCC. Accordingly, we would expect unstable behaviors because of many feed-back loops of the supply chains, as the literature suggests~\cite{Inoue2019}.

\section{Methods}

\subsection{Model}
\label{ch:model}

We employ the dynamic agent-based model of Inoue and Todo~\cite{Inoue2019,Inoue2019b}, which is an extension of the model of Hallegatte~\cite{Hallegatte08}.
In the model, each firm utilises inputs delivered from its supplier firms to produce an output and delivers it to client firms and final consumers.
Firms in the same industry are assumed to produce the same output. Supply chains are fixed over time (in two senses): 
First, each firm uses a firm-specific set of input varieties and does not change the input set over time. Second, each firm is linked with fixed suppliers and clients and cannot be linked with any new firms over time, even after a supply chain disruption.
Because supply chains are flexibly changed in the long run,
our analysis is focusing on short-term changes in production. Furthermore, we assume that each firm keeps inventories of each input at a level randomly determined from a Poisson distribution. Following Inoue and Todo~\cite{Inoue2019}, parameter values are calibrated based on the case of the Great East Japan earthquake. (see Supplementary Information~\ref{ch:appmodel} for details).

When a lockdown is imposed in a region,
firms in the region decrease their production because of restrictions on operating hours, the number of workers in offices and workshops, and workers' geographic mobility. The effect of this production reduction is propagated to both upstream and downstream firms. On the one hand, the demand of the restricted firm for intermediates from its suppliers immediately declines, and thus, the suppliers have to shrink their own production. Since the demand of suppliers' suppliers also declines, the effect of the restriction propagates upstream. On the other hand, the supply of products from the restricted firm to its client firms declines. Then, as the first reaction, client firms use their inventories of inputs to maintain the current level of production. In addition, client firms can procure inputs from other suppliers with additional production capacity in the same industry as the firms with declining supply. However, as explained, the supply chains is fixed, and the connection with the alternate supplier must already be in place before the restriction is imposed. If the inventories and inputs from substitute suppliers are insufficient, clients have to reduce their production because of an input shortage. Accordingly, the effect of the restriction propagates downstream through supply chains. Compared with upstream propagation, downstream propagation is likely to be slower because of the presence of inventory and substitution of suppliers.

\subsection{Restrictions in Japan}
\label{sec:soe}

The national or prefecture governments of Japan cannot legally impose strict lockdowns.
Rather than implementing restrictions through legal enforcement or punishments,
the governments can only request that workplaces close and that people stay at home and engage in social distancing. However, under strong social pressure in Japan, people and businesses voluntarily restrict their activities to a large extent. Although the details of the restrictions in Japan will be explained in the following paragraphs, hereafter, we do not distinguish between restrictions and lockdowns, if there is no such necessity.

In Japan, restrictions are imposed to each prefecture under the state of emergency~\cite{Cabinet2020}. It was first declared on 7 April 2020 in seven prefectures, Tokyo, Osaka, Fukuoka and their neighbouring prefectures, because these jurisdictions are metropolitan areas and thus showed a large number of confirmed COVID-19 cases. On 16 April, the state of emergency was expanded to all 47 prefectures. Then, it was lifted for 39 prefectures on 14 May, for an additional three on 21 May, and finally, for the remaining five prefectures on 25 May. 

A general implication of lockdowns is that all but substantial industries and activities are forced to halt operations. 
On the other hand, although the Japanese government declared a state of emergency,
how the restrictions were imposed was decided by each prefecture government. 

This study's purpose is to analyse each prefecture's restriction. As explained earlier, restrictions are captured in the model by the production capacity reduction. However, we cannot observe the extent to which each firm reduced its production capacity by obeying governmental requests or the rate of reduction in production capacity for each sector.
Therefore, it is necessary to incorporate some benchmark reduction level.
The benchmark proposed in the literature~\cite{Guan2020} is
that the rate of reduction in a sector is determined
by the degree of exposure to the virus
and the share of workers who cannot work at home given by~\cite{Bonadio2020}.
For example, in lifeline/essential sectors such as the utilities, health, and transport sectors, the rate of reduction is assumed to be zero; in other words, production capacity in these sectors does not change on lockdown. In sectors in which it is assumed that exposure to the virus is low (50\%) and 47.5\% of workers can work at home, such as the wholesale and retail sectors, the rate of reduction is 26.25\% ($=0.5 \times (1-0.475)$). Sectors with ordinary exposure (100\%) and a lower share of workers working at home (26.8\%), such as the iron and other metal product sectors, reduce production capacity by 26.8\% ($=1.0 \times (1-0.268)$). See Supplementary Information~\ref{ch:indmulti} for the rate of reduction in each sector.

Although the benchmark reduction rate from the literature offers a plausible value for any country, there must be some difference arising from the fact that Japan has different industrial structures or customs.
Inoue et al.~\cite{Inoue2020b} estimate this difference and find that
the model estimation with a weighted reduction rate of 32.3\%
best fits the actual losses observed in Japan. In this study, we follow their work and use the weight and worldwide reduction rate as the adjusted reduction rate shown in Table~\ref{tbl:mul} in the Supplementary Information.

\subsection{Simulations}
\label{sec:methodsim}

\paragraph*{Effect of one prefecture restriction}

We examine how a restriction in one prefecture affects production in another prefecture. For this purpose, we experiment with a restriction in one prefecture with the adjusted reduction rates and see how the effect propagates to other prefectures.
We evaluate different types of restrictions that vary in terms of the industrial sectors covered and the duration. For the industrial sectors, we assess four different coverage levels: (1) the accommodation and leisure sectors only; (2) the restaurant sector plus (1); (3) the retail sector plus (2); (4) all sectors. There is no decisive criterion on which of the four levels to use. However, the global lockdowns have displayed varying levels of coverage based on the level of urgency and amount of contact with final consumers in different sectors. Therefore, observing the different sector coverage levels can be informative. 
As the restriction duration, we test lengths of 1, 2, 3, and 4 weeks. 
As the experiment sets,
since Japan has 47 prefectures,
we have $47$ (prefectures) $\times$ $4$ (sector coverage levels) $\times$ $4$ (duration) sets of simulations.
Each set has 30 Monte Carlo runs.

For the duration of the restriction in a prefecture, all firms in the prefecture and the chosen sectors see a loss of production capacity. After this period, the production capacity immediately returns to its prerestriction volume. Moreover, we assume that inventories do not decay over time. Therefore, inventories stocked before the restriction can be fully utilized after the restriction is lifted. 

\paragraph*{Effect of two-prefecture coordination of restriction}

In practice, one prefecture often imposes a restriction when other prefectures do.
If the impact of concurrent operation, i.e., simultaneous imposition of restrictions in two or more prefectures, is smaller than that of asynchronous operations, i.e., imposition of restrictions two or more prefectures with no time overlap, it reveals the need for coordination.
We check all 47 $\times$ 46 $/$ 2 sets of two-prefecture combinations
and estimate the effects of asynchronous and concurrent restriction.
In this case, we evaluate restrictions on all industrial sectors only,
but the durations are varied at 1, 2, 3, and 4 weeks.
Each set has 30 Monte Carlo runs.

\paragraph*{Effect of nationwide coordination of restrictions}

As an extension of two-prefecture coordination scenario,
we analyse the scenario with coordination among all prefectures.
First, as in the two-prefectures case, we evaluate the scenario with concurrent restrictions across the 47 prefectures.
Regrading the asynchronous case, on the other hand, in practice,
it is probably not plausible to consider the 47 prefectures
imposing restrictions without no time ovelap.
Therefore, as the simulation to contrast with the concurrent simulation,
we randomly set the starts of the four-week restrictions in the 47 prefectures across the period of three months covering the winter season.
In the random selection of the duration for each prefecture, at least one prefecture imposes its restriction from the beginning (the first day).
We call this simulation the asynchronous case.
In this simulation, we evaluate a restriction on all sectors for four weeks only.
Each simulation has 100 Monte Carlo runs.

\section{Results}

\subsection{Effect of a one-prefecture restriction}
\label{ch:resultone}

\begin{figure}[tb]
\centering
\includegraphics[width=\linewidth]{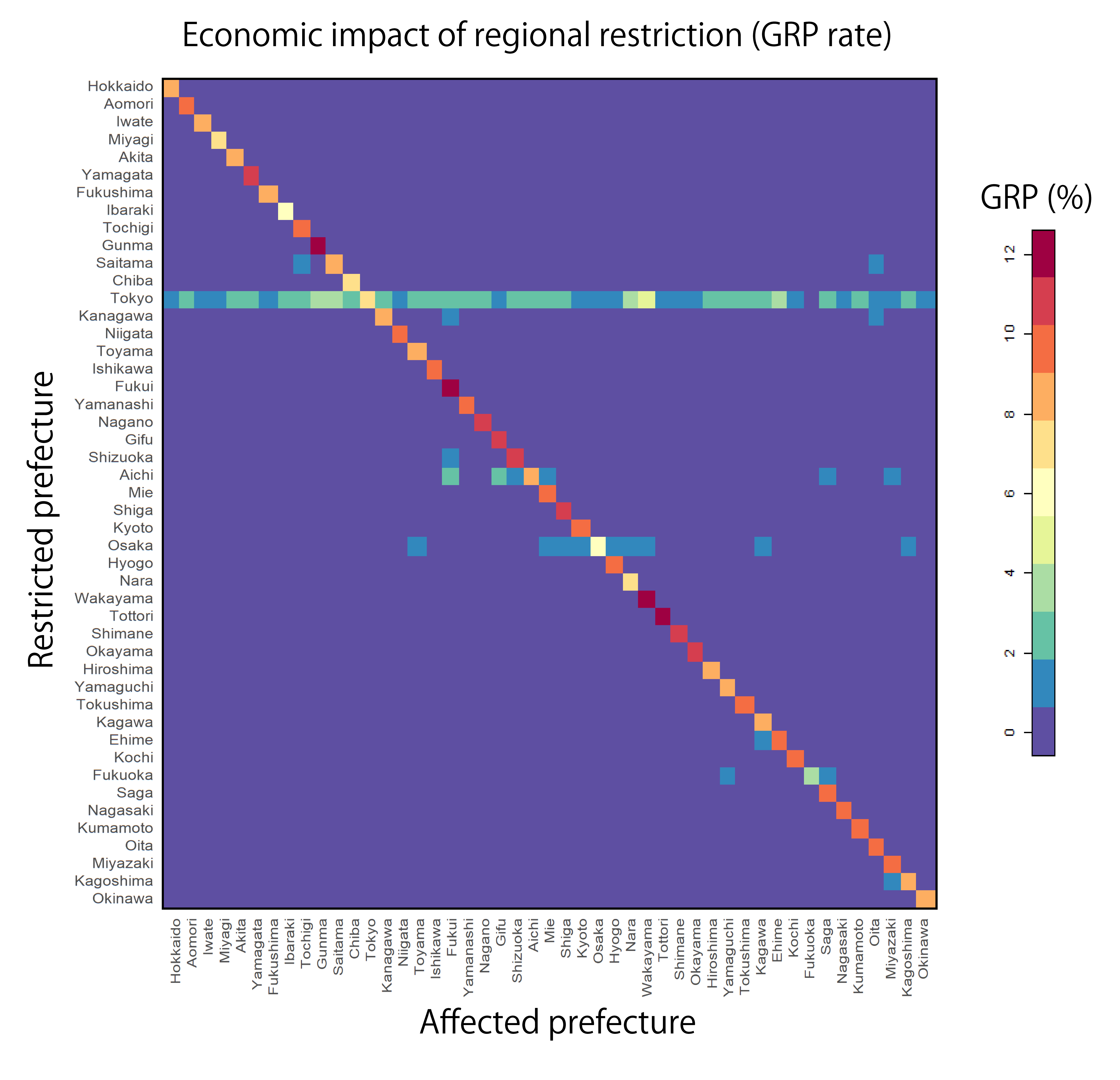}
\caption{Economic impact of a regional restriction. Each row shows a restriction in that prefecture. Each element shows an economic loss caused by the restriction in the prefecture. The magnitude of the loss of gross regional product (GRP) is shown by colors. The restriction in this figure covers all sectors and last four weeks.}
\label{fig:heatmap}
\end{figure}

Figure \ref{fig:heatmap} shows how a restriction in one prefecture affects production in other prefectures. The row side shows the prefectures on which the restriction is imposed.
The column side shows the prefectures affected by the restrictions. Therefore, the effect of each prefecture's restriction on other prefectures is shown on a horizontal line. The effect is calculated as the gross regional product (GRP) loss rate in a prefecture, which is shown by colors in the figure. Figure \ref{fig:heatmap} is for the case with an all-sector four-week lockdown. See Supplementary Information \ref{ch:onesi} for other cases.

Naturally, we see a diagonal line of GRP losses because the prefecture with the restriction experiences a direct loss.
Cells outside the diagonal line display the effect on prefectures with an indirect loss.
Since most of the cells are blue (close to 0\% GRP loss), it can be said that in the case of a one-prefecture restriction, the propagation effect is negligible. However, some prefectures have marked indirect effect.
A restriction in Tokyo has a very large indirect effect on other prefectures. Aichi and Osaka also have recognizable indirect effect, and a few other prefectures have some effects.

The total direct and indirect GRP losses (i.e., the GDP loss) are positively correlated with the GRP of the prefecture with the restriction. However, restrictions in some prefectures such as Kanagawa and Saitama impose smaller GDP losses than in other prefectures with lower GRPs such as Osaka and Hyogo. (See Figure \ref{fig:grpvs} in the Supplementary Information. Note that the lines have log scales.) This result means that the GRP of the prefecture cannot directly measure the indirect loss. This result shows that estimations considering actual supply chains are useful.

We also compare the effect of different lockdown durations.
(See Figures \ref{fig:7heat} and ~\ref{fig:28heat} in the Supplementary Information.) 
First, in terms of GRP loss, we can observe from colored dots in these figures, that there is not much difference between the one- and four-week simulations. This result is somewhat unexpected simply because if a restriction lasts longer, the damage should presumably be larger. Indeed, simulations for Tokyo in another study show that a four-week lockdown results in a larger losses than a one-week lockdown~\cite{Inoue2020}. One interpretation of this difference relates to substitutability.
The general implications of lockdown involve a very strong capacity loss.
On the other hand, the restrictions considered in this study are milder than a full lockdown.
As explained in Section \ref{ch:model}, substitution can be considered to affect intermediate goods. If supply disruption is too strong to be compensated by other suppliers, the downward pressure on production in the supplied firm could be large. On the other hand, if the disruption is small, it can be remedied by substitution from other suppliers.
Therefore, we presume that capacity reduction and GDP losses have a nonliner relationship. In other words, there is a tipping point at which the supply chain can absorb the negative shock through substitution.

Lastly, we compare the effect of different levels of sector coverage
(See Figures \ref{fig:7heat} and ~\ref{fig:28heat} in the Supplementary Information.) 
The sector coverage makes a difference, especially in the case of sector restriction levels (1) and (2).
These levels are (1) the accommodation and leisure sectors and (2) the restaurant sector plus (1).
Since those services are associated with spread of the infection,
governments tended to impose restrictions on those sectors.
Of course, as the sector coverage widens,
the GRP losses tend to become large.
However, we can see the marked gap between levels (1) and (2). This means
if a government closes the restaurant sector, the associated loss is its loss is comparable to wider sectoral restrictions.



\subsection{Effects of two-prefecture coordination of restrictions}

\begin{figure}[tb]
\centering
\includegraphics[width=\linewidth]{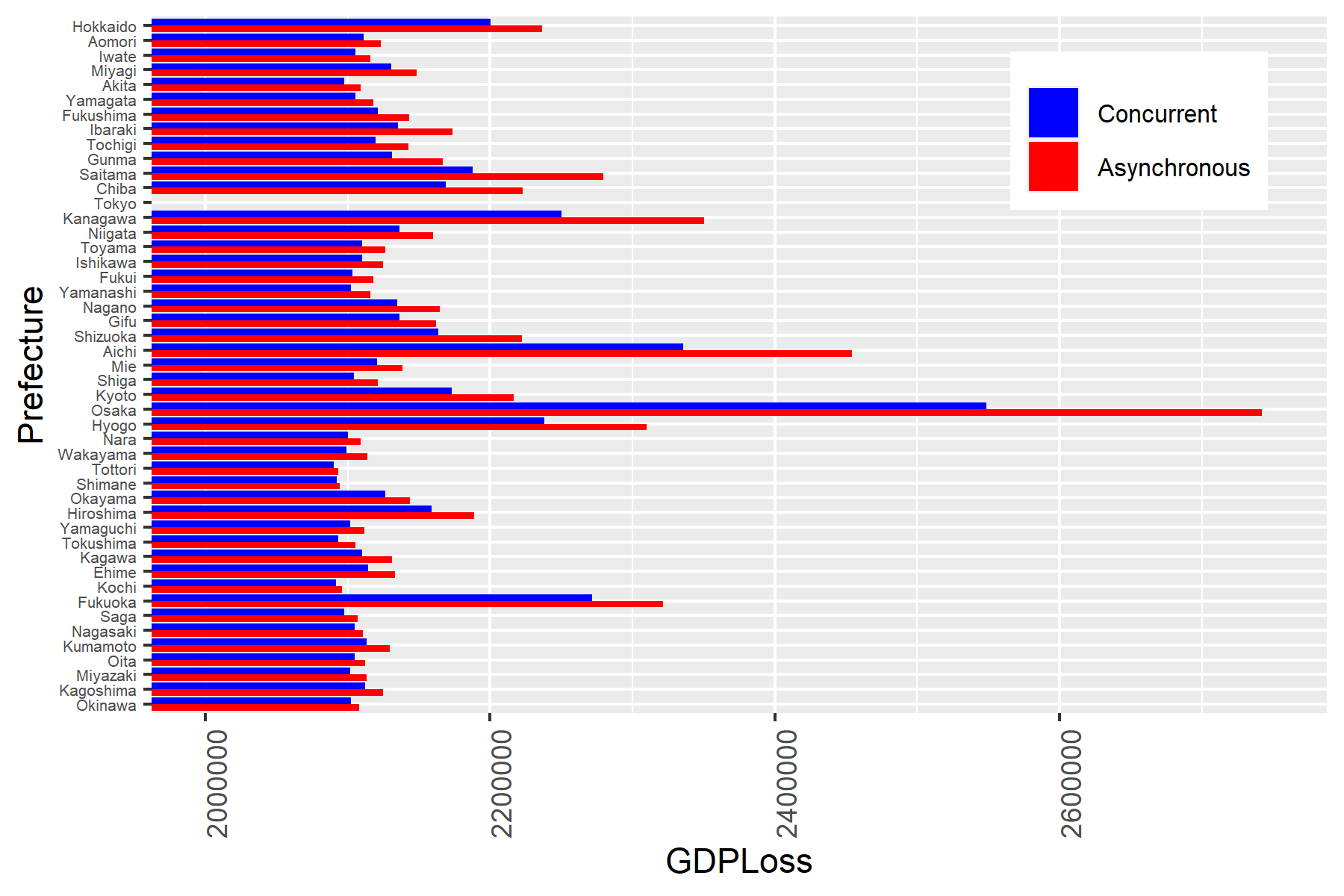}
\caption{Comparisons of asynchronous and concurrent simulations for restrictions in a given prefecture and Tokyo. The restrictions are imposed on the two prefectures in each simulation, with the prefectures with restrictions indicated on the vertical axis. The restrictions are imposed on all sectors for four weeks. The blue bars show the simulation restuls for asynchronous case, that is, the GDP loss from the two independent simulations, whereas the red bars show the GDP loss in the concurrent case. The bars are the average of 30 Monte Carlo simulations.}
\label{fig:congestion}
\end{figure}

The two-prefecture restriction scenario is tested to see whether 
the impacts of the concurrent and the asynchronous operations are different.
The restrictions are imposed on all sectors in this simulation.
We check all 47 $\times$ 46 $/$ 2 sets of two-prefecture combinations
and the four durations (1, 2, 3, and 4 weeks),
although it is not possible to show all the results. Therefore,
we highlight some important results with new findings.

Figure \ref{fig:congestion} shows combinations with a selected
prefecture and Tokyo.
Since Tokyo has a dominant effect on other prefectures, as shown in Section \ref{ch:resultone}, we highlight the combinations that include it.
The results in Figure \ref{fig:congestion} are for restrictions imposed on all sectors for four weeks. The ``concurrent'' label indicates that the two prefectures restrictions, for example, in Tokyo and Hokkaido, are imposed simultaneously for four weeks. The corresponding result is the second top red bar in Figure \ref{fig:congestion}.
The bar shows the GDP loss (entire supply chains loss).
On the other hand, the ``asynchronous'' label indicates the sum of the independent simulation, i.e., we test the scenarios with restrictions in Tokyo and Hokkaido imposed independently.
Then, we show the sum of the GDP losses from the independent simulations in the top bar in Figure \ref{fig:congestion}.

First, for any combinations of two prefectures, the GDP losses in the asynchronous case are larger than those in the concurrent case.
One reason for this result involves the demand and supply reductions.
Suppose that supplier A in prefecture X cuts its production and its demand from and supply to other firms.
If this occurs under an asynchronous restriction,
the direct loss is propagated to other firms.
However, suppose that there is another restriction to another prefecture Y, and let us consider client B in prefecture X of supplier A.
Since client B also reduces its production, its demand and supply decline.
The supply from A to B corresponds to the demand from B to A.
Therefore, both declines occur in both directions. Simply put,
they overlap.
However, if the restriction is imposed on only one prefecture, this overlap does not occur. Therefore, in the asynchronous case, the propagation is double that in the concurrent case.

Because of the simultaneousness that we explained above,
one may think that the result and the finding are not of importance.
However, this is not the case.
First, if the two prefectures do not have connections between them, the mitigation of overlaps is not likely to happen. In fact, there are several combinations of two prefectures without connections (They are not included in combinations of a prefecture and Tokyo. Tokyo are connected with all the other prefectures).
Even in such cases, we observe concurrent cases have smaller effect.
We presume that this effect comes from simultaneousness on indirect connections.
The example of simultaneousness in the last paragraph indicates a direct connection in two prefectures. However, even if two prefectures do not have direct connections, two prefectures are connected indirectly through firms in the third prefecture. Since this indirect simultaneousness would be smaller than the direct one, this simulation shows us that even in such cases, there is the advantage of the concurrent case.
In addition, as indicated in Section \ref{ch:resultone}, because of substitutability, if the negative shock is small, it may be absorbed by other supply chain partners. In other words, the size of the negative shock may be superlinearly associated with the GDP loss.
Therefore, the result shown in Figure \ref{fig:congestion}, i.e., that asynchronous restrictions mostly result in larger GDP losses than concurrent ones, is not obvious, and we uncover it only through the simulation.
Indeed, the combinations of the Toyama and Saga prefectures and the Wakayama and Kumamato prefectures show an inverse relationship; i.e., the concurrent restrictions imposes larger GDP losses than the asynchronous restrictions, although the gap is very small and is not statistically significant in the Wilcoxon rank sum test.
As a final analysis, in this restriction setting, the costs of concurrent restrictions are generally smaller than those of asynchronous restrictions.


Although Figure \ref{fig:congestion} highlights the combinations with Tokyo,
there are 46 other figures corresponding to Figure \ref{fig:congestion}.
Since it is not appropriate to show them all,
we put the figures for Aichi, Osaka, and Tottori (the prefecture with the smallest GRP) in the Supplementary Information \ref{ch:twosi}.
Through the results for the 47 prefectures and their combinations, 
we uncover another finding.
As shown in Figure \ref{fig:grpvs},
we can see the rank of the GDP losses from the imposition of a one-prefecture restriction.
A sum of two GDP losses is the result of the simulations of two asynchronous prefecture restrictions.
Therefore, the rank order of the GDP losses is the same as the rank order of the simulation results in the asynchronous cases.
For example, in the combinations of the different prefectures with Tokyo, the largest loss in the asynchronous case corresponds to Osaka, and the second largest to Aichi. This order is the same as the rank of the outcomes under the one-prefecture restriction.
On the other hand, the order of the concurrent simulation results does not completely mirror the order of the one-prefecture restriction simulation results.
For example, in Figure \ref{fig:congestion}, the order of the concurrent simulation outcomes displays anomalies for Hokkaido and Saitama. The isolated loss of Saitama is larger than that of Hokkaido.
However, the concurrent losses have the opposite order.
This anomaly can again be explained by the cancelling-out effect described above.
Saitama is located next to Tokyo. Therefore, they have strong supply chain connections.
Hence, it seems that the cancelling-out effect is more pronounced in Saitama  than in Hokkaido.
Since there are many examples of such anomalies, we can find a mechanism behind this phenomenon, but we leave this issue as the future work.


The results of the simulations with 1-, 2-, and 3-week restriction durations reveal the same findings as those obtained from four-week simulations.

\subsection{Effect of nationwide coordination of restrictions}

\begin{figure}[tb]
\centering
\includegraphics[width=0.8\linewidth]{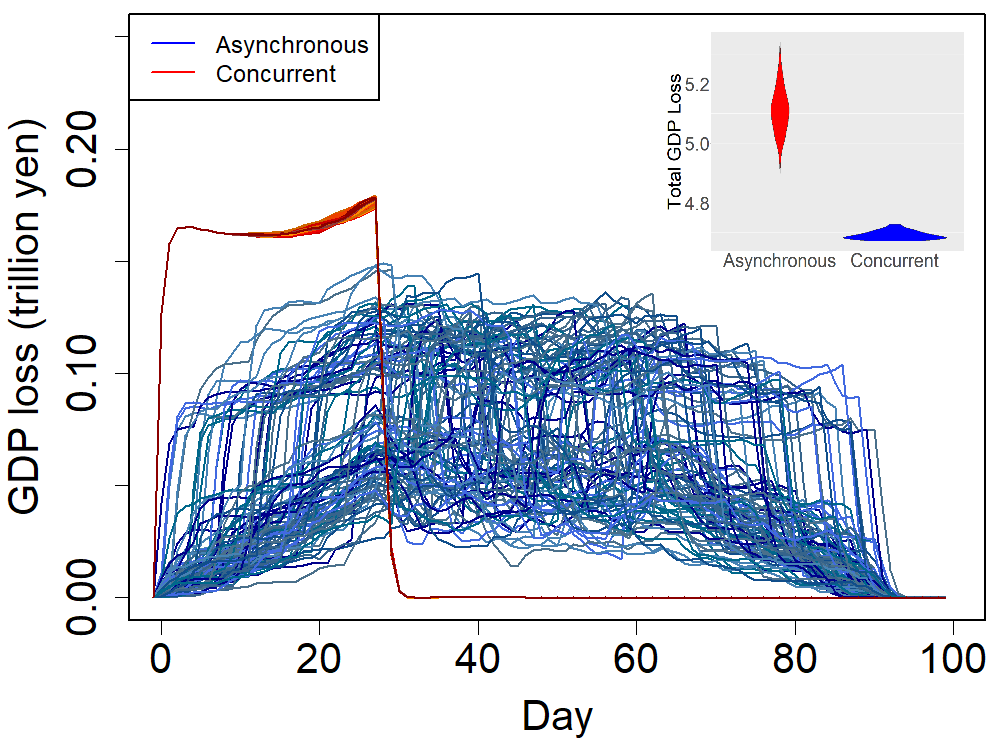}
\caption{Temporal comparisons of concurrent and asynchronous simulations for all prefecture restrictions. The vertical axis shows GDP. The horizontal axis shows days. The blue lines show the asynchronous restrictions in the prefectures with the four-week restriction timings randomly chosen from a range of three months. The red lines show four-week concurrent restrictions for all prefectures. One hundred simulations with different random seeds are shown in the asynchronous and concurrent simulations. The inset shows the violin plot for total GDP loss of the both settings.}
\label{fig:congestionAll}
\end{figure}

The restrictions with the combinations of two prefectures in the previous section are only one of possible scenario configurations.
Another possible and important scenario is the nationwide imposition of restrictions.
In this scenario, as in the concurrent case,
all 47 prefectures implement restrictions simultaneously.
In addition, as in the asynchronous case,
we set the duration of 47 prefectures' restrictions to three months with uniform randomness.
However, at least one prefecture imposes its restriction from the beginning of the three months.
The restrictions cover all sectors and last four weeks.

First, the GDP losses under the concurrent lockdowns are significantly smaller than those under the asynchronous lockdowns,
which is confirmed by the Wilcoxon rank sum test. The $p$ value is less than $10^{-10}$.
The average GDP losses in these cases are, respectively,
4.69 trillion yen or 43.8 billion U.S. dollars
and 5.11 trillion yen or 47.6 billion U.S. dollars.
Therefore, in principle, restrictions should be imposed simultaneously where possible.

Figure \ref{fig:congestionAll} shows
the temporal development of the simulations.
The concurrent restrictions naturally cause
intense GDP losses.
The asynchronous restrictions show dome-like shapes
and two-level stages, with daily GDP losses low in one stage,
but high in the other stage and no intermediate stages.
The cause of this result is the dominant effect of Tokyo.
Figure \ref{fig:congestionAllEx} in the Supplementary Information shows
5 out of 100 samples for the asynchronous case.
We see that the asynchronous lines remain in the high-level stage for four weeks.
The four weeks are the period when Tokyo is under the restriction.


\section{Discussion and Conclusion}

Our simulations show how the coordination of lockdowns affects the associated economic losses.
Our results and their implications are as follows.

First, GDP losses cannot be precisely predicted by the GRP of the locked-down regions. 
Lockdowns in some regions with small GRPs can cause large GDP losses.
In addition, since the losses are moderated by substitution of intermediate goods,
the initial shocks can be absorbed
but if the initial shocks exceed a certain
threshold, GDP losses can expand greatly,
which is observable in the case of different levels of sector coverage and lockdown durations in our simulations.

Second, coordinated, i.e., simultaneous, lockdowns generally cause lower GDP losses than
uncoordinated lockdowns. To show this fact, we evaluate completely simultaneous and completely independent lockdowns in two prefectures.
The difference between the coordinated and the uncoordinated cases presumably comes from
the cancelling-out effect.
Therefore, if the regions under lockdown are not complementary in terms of their supply chains,
the gap between the two cases may be small.

Finally,
a comparison of coordinated and uncoordinated lockdowns that cover entire supply chains reveals that the former are
significantly better.
To construct a plausible comparison, we set the uncoordinated lockdowns be imposed over three months at uniformly random times.
One possible explanation for this result is that
the coordinated lockdowns maximally leverage the cancelling-out effect.


The results of this study indicate
the necessity of policy coordination among regions
in regards to when governments should impose lockdowns.
To focus on firm-level supply chains, we limit our study to firms in Japan.
However, it in important to consider whether
the results of this study hold in international supply chains
and lockdowns imposed across them.
Indeed, the international propagation of negative shocks
has already been empirically observed
in the context of natural disasters.
Therefore, we presume that the effect of coordination of lockdowns discussed in this study
may hold for international coordination of lockdowns.

\section*{Acknowledgment}
This research used the computational resources of the supercomputer Fugaku (the evaluation environment in the trial phase) provided by the RIKEN Center for Computational Science. OACIS~\cite{murase2017open} and CARAVAN~\cite{murase2018caravan} were used for the simulations in this study.
This research was conducted as part of a project entitled ``Research on relationships between economic and social networks and globalization'' undertaken at the Research Institute of Economy, Trade, and Industry (RIETI). 
The authors are grateful for the financial support of JSPS Kakenhi Grant Nos. JP18K04615 and JP18H03642. 

\bibliographystyle{unsrt}
\bibliography{main}

\newpage

\appendix

\noindent{\huge {\bf Supplementary Information}}

\renewcommand\thefigure{\thesection.\arabic{figure}} 
\renewcommand\thetable{\thesection.\arabic{table}} 

\section{Data}
\label{ch:data}
\setcounter{figure}{0} 
\setcounter{table}{0} 

\subsection{Supply chains}
\label{ch:supch}

In the TSR data, the maximum number of suppliers and clients reported by each firm is 24. However, when we consider supplier-client relations running in the opposite direction, each firm can have more than 24 suppliers or 24 clients. Since the TSR data include the address of the headquarters of each firm, we can obtain the longitude and latitude of each headquarters using the geocoding service provided by the Center for Spatial Information Science at the University of Tokyo.

The TSR data do not include the volume of each transaction between two firms. Therefore, we estimate it. First, we divide each supplier's sales among its clients in proportion to the clients' sales to create a tentative sales value. Second, we refer to the IO table of Japan in 2015~\cite{METI15} to transform the tentative values so that the total volume corresponds to GDP. Specifically, we aggregate the tentative values at the firm-pair level to create the total sales for each pair of industrial sectors. We then divide the total sales for each industrial sector pair by the transaction values for the corresponding pair in the IO tables. The ratio is then used to adjust the transaction values between firms. The final consumption of each industrial sector is assigned to all firms in the sector using their sales as weights. 

\subsection{Prefectures in Japan}
\label{ch:prefmap}

\begin{figure}[htb]
\centering
\includegraphics[width=0.8\linewidth]{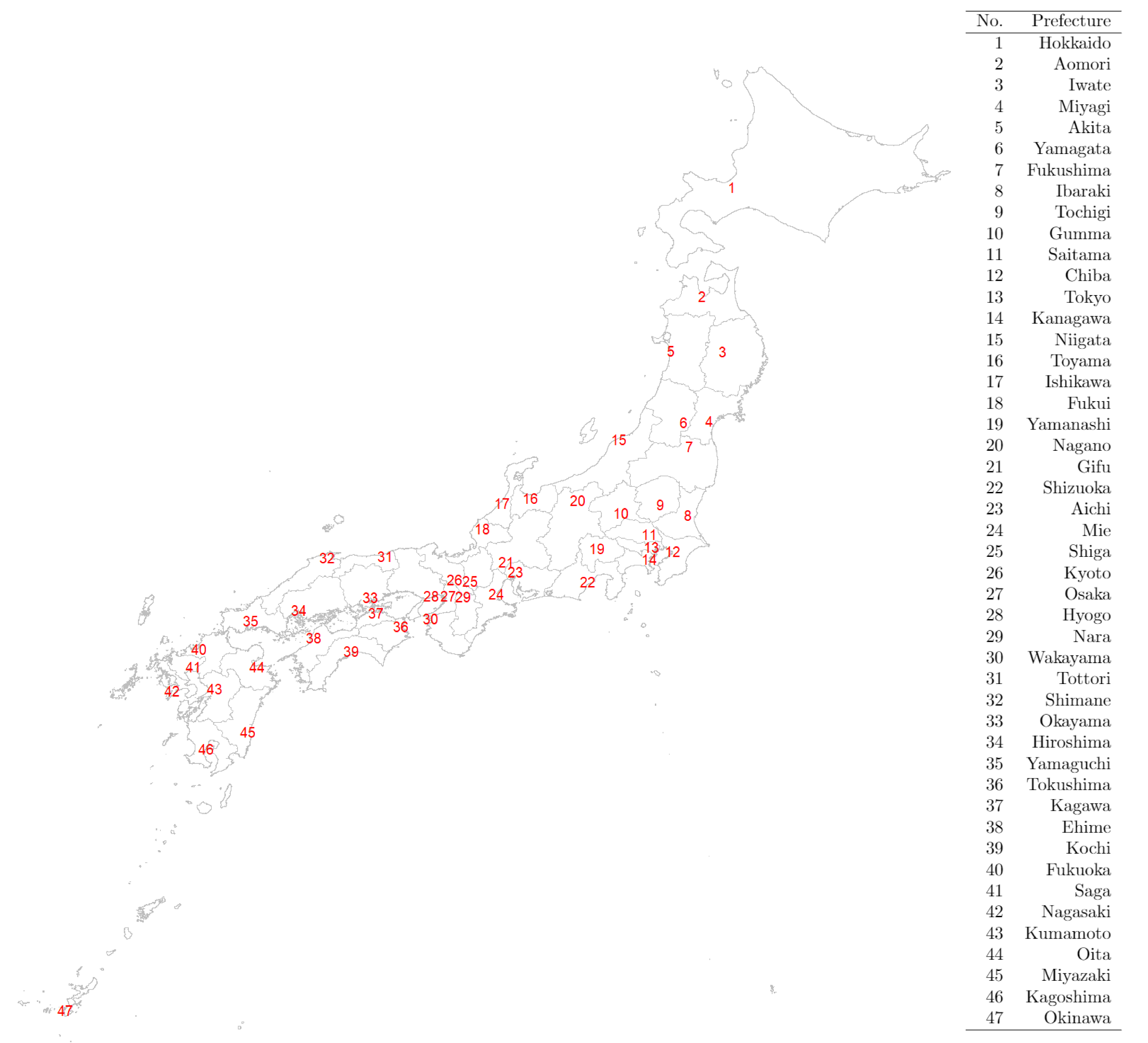}
\caption{Prefecture locations and codes. The number on the map is the JIS code of each prefecture shown in the table on the right. 
}
\label{fig:prefInfo}
\end{figure}

Since this study uses prefectures as the regional unit,
the geographic information of Japanese prefectures is necessary.
Figure~\ref{fig:prefInfo} shows the locations of the prefectures with the Japan Industrial Standard (JIS) codes and names of the 47 prefectures.
In the main text, the prefecture order shown in this figure is used.

\clearpage

\section{Methods}
\subsection{Model}
\label{ch:appmodel}
\setcounter{figure}{0} 
\setcounter{table}{0} 

We use the model of Inoue and Todo~\cite{Inoue2019, Inoue2019b}, an extension of agent-based models simulating the propagation of natural disaster shocks through supply chains~\cite{Hallegatte08}.
A brief explanation of the model is as follows: firms utilize various intermediates as inputs and provide a sector-specific product to client firms and final consumers. Firms have an inventory of intermediates from suppliers. However, they have no inventory of completed product, and the completed product is immediately delivered to clients. In addition, there are no price and no market mechanisms.

In the initial state before an economic shock, the daily trade volume from supplier $j$ to client $i$ is denoted by $A_{i,j}$, and the daily trade volume from firm $i$ to final consumers is denoted by $C_i$. Then, the initial production of firm $i$ in a day is given by 
\begin{equation}
P_{\mbox{ini}i}=\Sigma_j{A_{j,i}}+C_i.
\label{eq:p}
\end{equation}
On day $t$ after the initial state, the demand from the previous day for firm $i$'s product is $D_i^* (t-1)$. The firm thus makes orders to each supplier $j$
so that the amount of its product of supplier $j$ can meet this demand; therefore $A_{i,j}{D_i^*(t-1)}/{P_{\mbox{ini}i}}$. We assume that firm $i$ has an inventory of the intermediate goods produced by firm $j$ on day $t$, $S_{i,j} (t)$, and aims to restore this inventory to a level equal to a given number of days $n_i$ of the utilization of supplier $j$'s product. The constant $n_i$ is assumed to be Poisson distributed, where its mean is $n$, which is a parameter. In addition, $n_i$ does not take a number smaller than 4. We perform experiments to find the smallest number for the minimal inventory size that does not cause a bullwhip effect (a fluctuation of production levels). When the actual inventory is smaller than its target, firm $i$ increases its inventory gradually by $1/\tau$ of the gap so that it reaches the target in $\tau$ days, where $\tau$ is assumed to be 6 to follow the original model~\cite{Hallegatte08}. Therefore, firm $i$'s order to its supplier $j$ on day $t$, denoted by $O_{i,j}(t)$, is given by 
\begin{equation}
O_{i,j}(t)=A_{i,j}\frac{D_i^*(t-1)}{P_{\mbox{ini}i}}+\frac{1}{\tau}\left[n_i A_{i,j}-S_{i,j}(t)\right],
\label{eq:o}
\end{equation}
where the inventory gap is in brackets. Accordingly, total demand for the product of supplier $i$ on day $t$, $D_i(t)$, is given by the sum of final demand from final consumers and total orders from customers:
\begin{equation}
D_i(t)=\Sigma_jO_{j,i}(t)+C_i.
\end{equation}

Now, suppose that an economic shock hits the economy on day 0 and that firm $i$ is directly affected. Subsequently, proportion $\delta_{i}(t)$ of the production capital of firm $i$ malfunctions. In this study, $\delta_{i}$ is determined by the sector and prefecture to which firm $i$ belongs, and the duration of the imposition of the restriction. Hence, the production capacity of firm $i$, defined as its maximum production assuming no supply shortages, $P_{\mbox{cap}i}(t)$, is given by
\begin{equation}
P_{\mbox{cap}i}(t)=P_{\mbox{ini}i}(1-\delta_i(t)).
\end{equation}
The production of firm $i$ might also be limited by the shortage of supplies. Because we assume that firms in the same sector produce the same product, the shortage of supplies suffered by firm $j$ in sector $s$ can be compensated for by supplies from firm $k$ in the sector $s$. Firms cannot substitute new suppliers for affected suppliers after the disaster, as we assume fixed supply chains. Thus, the total inventory of the products delivered by firms in sector $s$ in firm $i$ on day $t$ is
\begin{equation}
S_{\mbox{tot}i,s}(t)=\Sigma_{j\in s}S_{i,j}(t).
\end{equation}
The initial consumption of products in sector $s$ of firm $i$ before the disaster is also defined for convenience as
\begin{equation}
A_{\mbox{tot}i,s}=\Sigma_{j\in s}A_{i,j}.
\end{equation}
The maximum possible production of firm $i$ limited by the product inventory of sector $s$ on day $t$, $P_{\mbox{pro}i,s}(t)$, is given by
\begin{equation}
P_{\mbox{pro}i,s}(t)=\frac{S_{\mbox{tot}i,s}(t)}{A_{\mbox{tot}i,s}}P_{\mbox{ini}i}.
\end{equation}
Then, we can determine the maximum production of firm $i$ on day $t$, considering its production capacity, $P_{\mbox{cap}i}(t)$, and its production constraints due to the shortage of supplies, $P_{\mbox{pro}i,s}(t)$:
\begin{equation}
P_{\mbox{max}i}(t)=\mbox{Min}\left(P_{\mbox{cap}i}(t), \mbox{Min}_{s}(P_{\mbox{pro}i,s}(t))\right). \label{eq:pconst}
\end{equation}
Therefore, the actual production of firm $i$ on day $t$ is given by
\begin{equation}
P_{\mbox{act}i}(t)=\mbox{Min}\left(P_{\mbox{max}i}(t), D_i(t)\right).
\label{eq:act}
\end{equation}

When demand for a firm's product is greater than its production capacity, the firm cannot completely satisfy the demand, as denoted by Equation (9). In this case, firms must ration their production to their customers. Following the previous work~\cite{Hallegatte08}, the rationing policy prioritises clients and final consumers who have relatively small orders to their initial (pre-shock) order, instead of being treated equally in the sense of absolute volume.

Suppose that firm $i$ has clients $j$ and a final consumer. Then, the order ratios, i.e., the order sizes relative to the initial state, from clients $j$ and the final consumer are denoted by $O^{rel}_{j,i}$ and $O^{rel}_{c}$, respectively. $O^{rel}_{\mbox{min}}=\mbox{Min}(O^{rel}_{j,i}, O^{rel}_{c})$ is calculated. If $r \leq (\sum_{j}{O^{rel}_{\mbox{min}}O_{j,i}}+O^{rel}_{\mbox{min}}C_i)$,
where $r$ is the remaining production of firm $i$,
$O^{rea}$ is calculated so that it satisfies $r=(\sum_{j}{O^{rea}O_{j,i}}+O^{rea}C_i)$.
Then we obtain $O_{j,i}^*=O^{rea}O_{j,i}+O^{sub}_{j,i}O_{j,i}$ and $C_i^*=O^{rea}C_{i}+O^{sub}_{c}C_{i}$,
where the realized order from firm $j$ to supplier $i$ is denoted by $O_{j,i}^* (t)$, and the realized order from a final consumer is $C_i^*$.
If $r > (\sum_{j}{O^{rel}_{\mbox{min}}O_{j,i}}+O^{rel}_{\mbox{min}}C_i)$,
we add $O^{rel}_{\mbox{min}}$ to $O^{sub}_{j,i}$ and $O^{sub}_{c}$,
where $O^{sub}_{j,i}$ and $O^{sub}_{c}$ are temporal variables used to calculate the realized order and are initially set to be zero.
Then, we subtract $(\sum_{j}{O^{rel}_{\mbox{min}}O_{j,i}}+O^{rel}_{\mbox{min}}C_i)$ from $r$ and remove the client or final consumer indicated by $O^{rel}_{\mbox{min}}$ from the calculation.
We repeat this process until $r$ is equal to zero.
More algorithmic and detailed explanations are found in~\cite{Inoue2020}. 

Under this rationing policy, total realized demand to firm $i$, $D_i^* (t)$, is given by
\begin{equation}
D_i^*(t)=\Sigma_jO_{i,j}^*(t)+C_i^*,
\end{equation}
where the realized order from firm $i$ to supplier $j$ is denoted by $O_{i,j}^*(t)$ and that from the final consumers by $C_i^*$. According to firms' production and procurement activities on day $t$, the inventory of firm $j$'s product in firm $i$ on day $t+1$ is updated to
\begin{equation}
S_{i,j}(t+1)=S_{i,j}(t)+O^{*}_{i,j}(t)-A_{i,j}\frac{P_{\mbox{act}i}(t-1)}{P_{\mbox{ini}i}}.
\end{equation}

Several caveats of this model and data should be mentioned. First, we assume that firms cannot find any new suppliers when facing a supply shortage from their current suppliers. 
Second, for simplicity, our model assumes that similar to manufacturing inputs, service sector inputs can be stored as inventory.  
Third, our model ignores changes in the prices of products and wages of labor incorporated in~\cite{Otto2017, Colon2019} and focuses on production dynamics during supply chain disruptions. 
Fourth, the TSR data report only the location of the headquarters of each firm, not the location of its branches. Because firms' headquarters are concentrated in Tokyo, production activities in Tokyo are most likely to be overvalued in our analysis. 
Fifth, because of data limitations, we ignore international supply chain links in our simulations. 
Finally, this study ignores the impacts of COVID-19 on human and firm behaviors in the post-COVID era. These behavioral changes may influence consumption and production, which are assumed to remain constant in this period.

\clearpage

\subsection{Sectoral differences in production capacity after restrictions}
\label{ch:indmulti}

There are no available firm- or sector-level data on production capacity (i.e., $P\mbox{cap}$ in the model) during the restriction in Japan. To obtain the production capacity of each firm, we assume that the rate of reduction in production capacity for each sector is given by the degree of the aggregated reduction because of exposure to the virus~\cite{Bonadio2020} multiplied by the share of workers who cannot work at home~\cite{Guan2020}. Both of the cited studies deal with the global situation in April and May in 2020. The rate of reduction because of exposure to the virus is determined by how many workers in the sector reduced their activities to avoid contact with others to prevent  infections. Because~\cite{Guan2020} defines the rate of reduction uniformly worldwide, we modify the rate for some sectors that clearly differ in the case of Japan. Table~\ref{tbl:mul} shows the reduction rate for each sector assumed in~\cite{Guan2020}.

Although it is assumed that the reduction rate is applicable to any country,
there must be a difference between the real rate of the country and the assumed reduction rate.
Inoue et al.~\cite{Inoue2020b} estimate this difference for Japan.
First, they estimate the GDP loss by using a real economic loss estimated by 
the Indices of All Industry Activities (IAA) for April and May of 2020.
The estimated loss is 7.52 trillion yen (71.2 billion U.S. dollars at 107 yen / dollar.) or 2.5\% of yearly GDP.
On the other hand, if we use the reduction rate derived as explained above,
the estimated GDP loss is 35.0 trillion yen (327 billion U.S. dollars).
It can be presumed that the actual reduction rate is smaller than the worldwide reduction rate. Therefore, the rate is adjusted by multiplying the worldwide reduction rate by the same weight so that the simulation estimation fits the IAIA estimation. The weight of 32.3\% provides the closest fit to the IAIA estimation.
In this study, we use this weight and the worldwide reduction rate as the adjusted reduction rate shown in Table~\ref{tbl:mul}.

{\footnotesize
\begin{longtable}{| l | l | l | l | l | l | l |} 
\caption{Sector-specific rates of reduction in production capacity. Sectors are classified according to the JIS classification~\cite{MIC2013} at the two-digit level, except for industries 560, 561, and 569 for which we use three-digit codes to reflect current circumstances. The sector names are abbreviated. Table~\ref{tbl:abb} lists the sector descriptions and abbreviations.} \label{tbl:mul}\\
\hline
Code & Sector & Adjusted reduction & Reduction & Work-at & Exposure & Rationale \\
 & (abbreviated) & rate & rate & -home rate & level & \\
\hline
\hline
1 & AGR. & 1.40E-01 & 4.33E-01 & 1.34E-01 & 0.5 & Low exposure \\
2 & FRS. & 1.40E-01 & 4.33E-01 & 1.34E-01 & 0.5 & Low exposure \\
3 & FIS. & 1.40E-01 & 4.33E-01 & 1.34E-01 & 0.5 & Low exposure \\
4 & AQA. & 1.40E-01 & 4.33E-01 & 1.34E-01 & 0.5 & Low exposure \\
5 & MIN. & 2.06E-01 & 6.37E-01 & 3.63E-01 & 1 & Ordinary \\
6 & CNS.GEN. & 2.45E-01 & 7.58E-01 & 2.42E-01 & 1 & Ordinary \\
7 & CNS.SPC. & 2.45E-01 & 7.58E-01 & 2.42E-01 & 1 & Ordinary \\
8 & EQP. & 2.45E-01 & 7.58E-01 & 2.42E-01 & 1 & Ordinary \\
9 & MAN.FOD. & 2.45E-01 & 7.60E-01 & 2.40E-01 & 1 & Ordinary \\
10 & MAN.BEV. & 2.45E-01 & 7.60E-01 & 2.40E-01 & 1 & Ordinary \\
11 & MAN.TEX & 2.16E-01 & 6.68E-01 & 3.32E-01 & 1 & Ordinary \\
12 & MAN.LUM. & 2.48E-01 & 7.68E-01 & 2.32E-01 & 1 & Ordinary \\
13 & MAN.FUR. & 2.48E-01 & 7.68E-01 & 2.32E-01 & 1 & Ordinary \\
14 & MAN.PUL. & 2.18E-01 & 6.76E-01 & 3.24E-01 & 1 & Ordinary \\
15 & PRT. & 2.18E-01 & 6.76E-01 & 3.24E-01 & 1 & Ordinary \\
16 & MAN.CHM. & 1.71E-01 & 5.29E-01 & 4.71E-01 & 1 & Ordinary \\
17 & MAN.PET. & 2.10E-01 & 6.51E-01 & 3.49E-01 & 1 & Ordinary \\
18 & MAN.PLA. & 2.27E-01 & 7.04E-01 & 2.96E-01 & 1 & Ordinary \\
19 & MAN.RUB. & 2.27E-01 & 7.04E-01 & 2.96E-01 & 1 & Ordinary \\
20 & MAN.LET. & 2.16E-01 & 6.68E-01 & 3.32E-01 & 1 & Ordinary \\
21 & MAN.CER. & 2.29E-01 & 7.09E-01 & 2.91E-01 & 1 & Ordinary \\
22 & MAN.IRN. & 2.36E-01 & 7.32E-01 & 2.68E-01 & 1 & Ordinary \\
23 & MAN.NFM. & 2.36E-01 & 7.32E-01 & 2.68E-01 & 1 & Ordinary \\
24 & MAN.FBM. & 2.24E-01 & 6.95E-01 & 3.05E-01 & 1 & Ordinary \\
25 & MAN.GNM. & 1.95E-01 & 6.04E-01 & 3.96E-01 & 1 & Ordinary \\
26 & MAN.PRM. & 1.95E-01 & 6.04E-01 & 3.96E-01 & 1 & Ordinary \\
27 & MAN.BSM. & 1.95E-01 & 6.04E-01 & 3.96E-01 & 1 & Ordinary \\
28 & EPT. & 1.08E-01 & 3.33E-01 & 6.67E-01 & 1 & Ordinary \\
29 & MAN.ELM. & 1.87E-01 & 5.80E-01 & 4.20E-01 & 1 & Ordinary \\
30 & MAN.INF. & 1.08E-01 & 3.33E-01 & 6.67E-01 & 1 & Ordinary \\
31 & MAN.TRN. & 1.63E-01 & 5.04E-01 & 4.96E-01 & 1 & Ordinary \\
32 & MAN.MSC. & 2.28E-01 & 7.05E-01 & 2.95E-01 & 1 & Ordinary \\
33 & ELE. & 2.01E-02 & 6.23E-02 & 3.77E-01 & 0.1 & Lifeline \\
34 & GAS. & 2.01E-02 & 6.23E-02 & 3.77E-01 & 0.1 & Lifeline \\
35 & HET. & 2.01E-02 & 6.23E-02 & 3.77E-01 & 0.1 & Lifeline \\
36 & WTR. & 2.01E-02 & 6.23E-02 & 3.77E-01 & 0.1 & Lifeline \\
37 & COM. & 1.30E-02 & 4.01E-02 & 5.99E-01 & 0.1 & Lifeline \\
38 & BRD. & 6.20E-03 & 1.92E-02 & 8.08E-01 & 0.1 & Lifeline \\
39 & INF.SVC. & 3.13E-02 & 9.70E-02 & 9.03E-01 & 1 & Ordinary \\
40 & INT. & 1.30E-02 & 4.01E-02 & 5.99E-01 & 0.1 & Lifeline \\
41 & INF.DST. & 6.20E-02 & 1.92E-01 & 8.08E-01 & 1 & Ordinary \\
42 & RLW.TRP. & 2.26E-02 & 7.01E-02 & 2.99E-01 & 0.1 & Lifeline \\
43 & PAS.TRP. & 2.26E-02 & 7.01E-02 & 2.99E-01 & 0.1 & Lifeline \\
44 & FRE.TRP. & 2.26E-02 & 7.01E-02 & 2.99E-01 & 0.1 & Lifeline \\
45 & WTR.TRP. & 2.26E-02 & 7.01E-02 & 2.99E-01 & 0.1 & Lifeline \\
46 & AIR.TRP. & 2.26E-02 & 7.01E-02 & 2.99E-01 & 0.1 & Lifeline \\
47 & WRH. & 2.26E-02 & 7.01E-02 & 2.99E-01 & 0.1 & Lifeline \\
48 & SVC.TRP. & 2.26E-02 & 7.01E-02 & 2.99E-01 & 0.1 & Lifeline \\
49 & PST.SVC. & 2.26E-02 & 7.01E-02 & 2.99E-01 & 0.1 & Lifeline \\
50 & WHL.GEN. & 1.70E-01 & 5.25E-01 & 4.75E-01 & 1 & Ordinary \\
51 & WHL.TEX. & 1.70E-01 & 5.25E-01 & 4.75E-01 & 1 & Ordinary \\
52 & WHL.FOD. & 1.70E-01 & 5.25E-01 & 4.75E-01 & 1 & Ordinary \\
53 & WHL.MAT. & 1.70E-01 & 5.25E-01 & 4.75E-01 & 1 & Ordinary \\
54 & WHL.MCN. & 1.70E-01 & 5.25E-01 & 4.75E-01 & 1 & Ordinary \\
55 & WHL.MSC. & 1.70E-01 & 5.25E-01 & 4.75E-01 & 1 & Ordinary \\
560 & RTL.ADM. & 1.70E-01 & 5.25E-01 & 4.75E-01 & 1 & Ordinary \\
561 & RTL.DPT. & 1.70E-01 & 5.25E-01 & 4.75E-01 & 1 & Closed \\
569 & RTL.GNM. & 1.70E-02 & 5.25E-02 & 4.75E-01 & 0.1 & Lifeline \\
57 & RTL.GEN. & 1.70E-01 & 5.25E-01 & 4.75E-01 & 1 & Ordinary \\
58 & RTL.FOD. & 1.70E-01 & 5.25E-01 & 4.75E-01 & 1 & Ordinary \\
59 & RTL.MCN. & 1.70E-01 & 5.25E-01 & 4.75E-01 & 1 & Ordinary \\
60 & RTL.MSC. & 1.70E-01 & 5.25E-01 & 4.75E-01 & 1 & Ordinary \\
61 & RTL.NST. & 1.70E-01 & 5.25E-01 & 4.75E-01 & 1 & Ordinary \\
62 & FIN.BNK. & 6.91E-02 & 2.14E-01 & 7.86E-01 & 1 & Ordinary \\
63 & FIN.ORG. & 6.91E-02 & 2.14E-01 & 7.86E-01 & 1 & Ordinary \\
64 & FIN.LON. & 6.91E-02 & 2.14E-01 & 7.86E-01 & 1 & Ordinary \\
65 & FIN.TRN. & 6.91E-02 & 2.14E-01 & 7.86E-01 & 1 & Ordinary \\
66 & FIN.AUX. & 6.91E-02 & 2.14E-01 & 7.86E-01 & 1 & Ordinary \\
67 & INS. & 6.91E-02 & 2.14E-01 & 7.86E-01 & 1 & Ordinary \\
68 & RST.AGN. & 1.37E-01 & 4.23E-01 & 5.77E-01 & 1 & Ordinary \\
69 & RTS.LES. & 1.37E-01 & 4.23E-01 & 5.77E-01 & 1 & Ordinary \\
70 & RNT. & 1.17E-01 & 3.62E-01 & 6.38E-01 & 1 & Ordinary \\
71 & SCI. & 5.56E-02 & 1.72E-01 & 8.28E-01 & 1 & Ordinary \\
72 & SVC.PRF. & 1.17E-01 & 3.62E-01 & 6.38E-01 & 1 & Ordinary \\
73 & ADV. & 1.17E-01 & 3.62E-01 & 6.38E-01 & 1 & Ordinary \\
74 & SVC.TEC. & 1.17E-01 & 3.62E-01 & 6.38E-01 & 1 & Ordinary \\
75 & ACM. & 2.87E-01 & 8.89E-01 & 1.11E-01 & 1 & Closed \\
76 & EAT. & 2.87E-01 & 8.89E-01 & 1.11E-01 & 1 & Ordinary \\
77 & DEL. & 1.68E-02 & 5.21E-02 & 4.79E-01 & 0.1 & Lifeline \\
78 & LND. & 1.68E-01 & 5.21E-01 & 4.79E-01 & 1 & Ordinary \\
79 & SVC.PSN. & 1.68E-01 & 5.21E-01 & 4.79E-01 & 1 & Ordinary \\
80 & SVC.AMS. & 1.68E-01 & 5.21E-01 & 4.79E-01 & 1 & Closed \\
81 & SCH. & 2.78E-02 & 8.60E-02 & 8.28E-01 & 0.5 & Low exposure \\
82 & EDC. & 2.78E-02 & 8.60E-02 & 8.28E-01 & 0.5 & Low exposure \\
83 & MED. & 2.43E-02 & 7.53E-02 & 2.47E-01 & 0.1 & Lifeline \\
84 & HLT. & 0.00E+00 & 0.00E+00 & 2.47E-01 & 0 & Sustantial \\
85 & WEL. & 0.00E+00 & 0.00E+00 & 2.47E-01 & 0 & Sustantial \\
86 & PST.OFC. & 1.17E-02 & 3.62E-02 & 6.38E-01 & 0.1 & Lifeline \\
87 & CAS. & 5.85E-02 & 1.81E-01 & 6.38E-01 & 0.5 & Low exposure \\
88 & WAS. & 5.85E-02 & 1.81E-01 & 6.38E-01 & 0.5 & Low exposure \\
89 & SVC.AUT. & 5.85E-02 & 1.81E-01 & 6.38E-01 & 0.5 & Low exposure \\
90 & SVC.MCN. & 5.85E-02 & 1.81E-01 & 6.38E-01 & 0.5 & Low exposure \\
91 & SVC.EMP. & 5.85E-02 & 1.81E-01 & 6.38E-01 & 0.5 & Low exposure \\
92 & SVC.BUS. & 5.85E-02 & 1.81E-01 & 6.38E-01 & 0.5 & Low exposure \\
93 & PLT. & 5.85E-02 & 1.81E-01 & 6.38E-01 & 0.5 & Low exposure \\
94 & REL. & 5.85E-02 & 1.81E-01 & 6.38E-01 & 0.5 & Low exposure \\
95 & SVC.MSC. & 5.85E-02 & 1.81E-01 & 6.38E-01 & 0.5 & Low exposure \\
96 & GOV.INT. & 1.66E-02 & 5.15E-02 & 4.85E-01 & 0.1 & Lifeline \\
97 & NA & 1.66E-02 & 5.15E-02 & 4.85E-01 & 0.1 & Lifeline \\
98 & GOV.LOC. & 1.66E-02 & 5.15E-02 & 4.85E-01 & 0.1 & Lifeline \\
99 & NEC & 1.17E-01 & 3.62E-01 & 6.38E-01 & 1 & Ordinary \\

\hline
\end{longtable}
}

\begin{landscape}
{\footnotesize
\begin{longtable}{| l | l | l |} 
\caption{Sector classifications and abbreviations.} \label{tbl:abb}\\
\hline
Code & Description & Abbreviation \\
\hline
\hline
01 & AGRICULTURE & AGR. \\
02 & FORESTRY & FRS. \\
03 & FISHERIES, EXCEPT AQUACULTURE & FIS. \\
04 & AQUACULTURE & AQA. \\
05 & MINING AND QUARRYING OF STONE AND GRAVEL & MIN. \\
06 & CONSTRUCTION WORK, GENERAL INCLUDING PUBLIC AND PRIVATE CONSTRUCTION WORK & CNS.GEN. \\
07 & CONSTRUCTION WORK BY SPECIALIST CONTRACTOR, EXCEPT EQUIPMENT INSTALLATION WORK & CNS.SPC. \\
08 & EQUIPMENT INSTALLATION WORK & EQP. \\
09 & MANUFACTURE OF FOOD & MAN.FOD. \\
10 & MANUFACTURE OF BEVERAGES, TOBACCO AND FEED & MAN.BEV. \\
11 & MANUFACTURE OF TEXTILE PRODUCTS & MAN.TEX \\
12 & MANUFACTURE OF LUMBER AND WOOD PRODUCTS, EXCEPT FURNITURE & MAN.LUM. \\
13 & MANUFACTURE OF FURNITURE AND FIXTURES & MAN.FUR. \\
14 & MANUFACTURE OF PULP, PAPER AND PAPER PRODUCTS & MAN.PUL. \\
15 & PRINTING AND ALLIED INDUSTRIES & PRT. \\
16 & MANUFACTURE OF CHEMICAL AND ALLIED PRODUCTS & MAN.CHM. \\
17 & MANUFACTURE OF PETROLEUM AND COAL PRODUCTS & MAN.PET. \\
18 & MANUFACTURE OF PLASTIC PRODUCTS, EXCEPT OTHERWISE CLASSIFIED & MAN.PLA. \\
19 & MANUFACTURE OF RUBBER PRODUCTS & MAN.RUB. \\
20 & MANUFACTURE OF LEATHER TANNING, LEATHER PRODUCTS AND FUR SKINS & MAN.LET. \\
21 & MANUFACTURE OF CERAMIC, STONE AND CLAY PRODUCTS & MAN.CER. \\
22 & MANUFACTURE OF IRON AND STEEL & MAN.IRN. \\
23 & MANUFACTURE OF NON-FERROUS METALS AND PRODUCTS & MAN.NFM. \\
24 & MANUFACTURE OF FABRICATED METAL PRODUCTS & MAN.FBM. \\
25 & MANUFACTURE OF GENERAL-PURPOSE MACHINERY & MAN.GNM. \\
26 & MANUFACTURE OF PRODUCTION MACHINERY & MAN.PRM. \\
27 & MANUFACTURE OF BUSINESS ORIENTED MACHINERY & MAN.BSM. \\
28 & ELECTRONIC PARTS, DEVICES AND ELECTRONIC CIRCUITS & EPT. \\
29 & MANUFACTURE OF ELECTRICAL MACHINERY, EQUIPMENT AND SUPPLIES & MAN.ELM. \\
30 & MANUFACTURE OF INFORMATION AND COMMUNICATION ELECTRONICS EQUIPMENT & MAN.INF. \\
31 & MANUFACTURE OF TRANSPORTATION EQUIPMENT & MAN.TRN. \\
32 & MISCELLANEOUS MANUFACTURING INDUSTRIES & MAN.MSC. \\
33 & PRODUCTION, TRANSMISSION AND DISTRIBUTION OF ELECTRICITY & ELE. \\
34 & PRODUCTION AND DISTRIBUTION OF GAS & GAS. \\
35 & HEAT SUPPLY & HET. \\
36 & COLLECTION, PURIFICATION AND DISTRIBUTION OF WATER, AND SEWAGE COLLECTION, PROCESSING & WTR. \\
37 & COMMUNICATIONS & COM. \\
38 & BROADCASTING & BRD. \\
39 & INFORMATION SERVICES & INF.SVC. \\
40 & SERVICES INCIDENTAL TO INTERNET & INT. \\
41 & VIDEO PICTURE INFORMATION, SOUND INFORMATION, CHARACTER INFORMATION PRODUCTION AND DISTRIBUTION & INF.DST. \\
42 & RAILWAY TRANSPORT & RLW.TRP. \\
43 & ROAD PASSENGER TRANSPORT & PAS.TRP. \\
44 & ROAD FREIGHT TRANSPORT & FRE.TRP. \\
45 & WATER TRANSPORT & WTR.TRP. \\
46 & AIR TRANSPORT & AIR.TRP. \\
47 & WAREHOUSING & WRH. \\
48 & SERVICES INCIDENTAL TO TRANSPORT & SVC.TRP. \\
49 & POSTAL SERVICES, INCLUDING MAIL DELIVERY & PST.SVC. \\
50 & WHOLESALE TRADE, GENERAL MERCHANDISE & WHL.GEN. \\
51 & WHOLESALE TRADE (TEXTILE AND APPAREL) & WHL.TEX. \\
52 & WHOLESALE TRADE (FOOD AND BEVERAGES) & WHL.FOD. \\
53 & WHOLESALE TRADE (BUILDING MATERIALS, MINERALS AND METALS, ETC) & WHL.MAT. \\
54 & WHOLESALE TRADE (MACHINERY AND EQUIPMENT) & WHL.MCN. \\
55 & MISCELLANEOUS WHOLESALE TRADE & WHL.MSC. \\
560 & ESTABLISHMENTS ENGAGED IN ADMINISTRATIVE OR ANCILLARY ECONOMIC ACTIVITIES & RTL.ADM. \\
561 & DEPARTMENT STORES AND GENERAL MERCHANDISE SUPERMARKET & RTL.DPT. \\
569 & MISCELLANEOUS RETAIL TRADE, GENERAL MERCHANDISE  & RTL.GMN. \\
57 & RETAIL TRADE, GENERAL MERCHANDISE & RTL.GEN. \\
58 & RETAIL TRADE (FOOD AND BEVERAGE) & RTL.FOD. \\
59 & RETAIL TRADE (MACHINERY AND EQUIPMENT) & RTL.MCN. \\
60 & MISCELLANEOUS RETAIL TRADE & RTL.MSC. \\
61 & NONSTORE RETAILERS & RTL.NST. \\
62 & BANKING & FIN.BNK. \\
63 & FINANCIAL INSTITUTIONS FOR COOPERATIVE ORGANIZATIONS & FIN.ORG. \\
64 & NON-DEPOSIT MONEY CORPORATIONS, INCLUDING LENDING AND CREDIT CARD BUSINESS & FIN.LON. \\
65 & FINANCIAL PRODUCTS TRANSACTION DEALERS AND FUTURES COMMODITY TRANSACTION DEALERS & FIN.TRN. \\
66 & FINANCIAL AUXILIARIES & FIN.AUX. \\
67 & INSURANCE INSTITUTIONS, INCLUDING INSURANCE AGENTS, BROKERS AND SERVICES & INS. \\
68 & REAL ESTATE AGENCIES & RST.AGN. \\
69 & REAL ESTATE LESSORS AND MANAGERS & RTS.LES. \\
70 & GOODS RENTAL AND LEASING & RNT. \\
71 & SCIENTIFIC AND DEVELOPMENT RESEARCH INSTITUTES & SCI. \\
72 & PROFESSIONAL SERVICES, N.E.C. & SVC.PRF. \\
73 & ADVERTISING & ADV. \\
74 & TECHNICAL SERVICES, N.E.C. & SVC.TEC. \\
75 & ACCOMMODATION & ACM. \\
76 & EATING AND DRINKING PLACES & EAT. \\
77 & FOOD TAKE OUT AND DELIVERY SERVICES & DEL. \\
78 & LAUNDRY, BEAUTY AND BATH SERVICES & LND. \\
79 & MISCELLANEOUS LIVING-RELATED AND PERSONAL SERVICES & SVC.PSN. \\
80 & SERVICES FOR AMUSEMENT AND RECREATION & SVC.AMS. \\
81 & SCHOOL EDUCATION & SCH. \\
82 & MISCELLANEOUS EDUCATION, LEARNING SUPPORT & EDC. \\
83 & MEDICAL AND OTHER HEALTH SERVICE & MED. \\
84 & PUBLIC HEALTH AND HYGIENE & HLT. \\
85 & SOCIAL INSURANCE, SOCIAL WELFARE AND CARE SERVICES & WEL. \\
86 & POSTAL OFFICE & PST.OFC. \\
87 & COOPERATIVE ASSOCIATIONS, N.E.C. & CAS. \\
88 & WASTE DISPOSAL BUSINESS & WAS. \\
89 & AUTOMOBILE MAINTENANCE SERVICES & SVC.AUT. \\
90 & MACHINE, ETC. REPAIR SERVICES, EXCEPT OTHERWISE CLASSIFIED & SVC.MCN. \\
91 & EMPLOYMENT AND WORKER DISPATCHING SERVICES & SVC.EMP. \\
92 & MISCELLANEOUS BUSINESS SERVICES & SVC.BUS. \\
93 & POLITICAL, BUSINESS AND CULTURAL ORGANIZATIONS & PLT. \\
94 & RELIGION & REL. \\
95 & MISCELLANEOUS SERVICES & SVC.MSC. \\
96 & FOREIGN GOVERNMENTS AND INTERNATIONAL AGENCIES IN JAPAN & GOV.INT. \\
97 & NATIONAL GOVERNMENT SERVICES & GOV.NAT. \\
98 & LOCAL GOVERNMENT SERVICES & GOV.LOC. \\
99 & INDUSTRIES UNABLE TO CLASSIFY & NEC \\
\hline
\end{longtable}
}
\end{landscape}

\clearpage

\section{Results}

\subsection{Effects of a one-prefecture restriction}
\label{ch:onesi}
\setcounter{figure}{0} 
\setcounter{table}{0} 

\begin{figure}[htb]
\centering
\includegraphics[width=\linewidth]{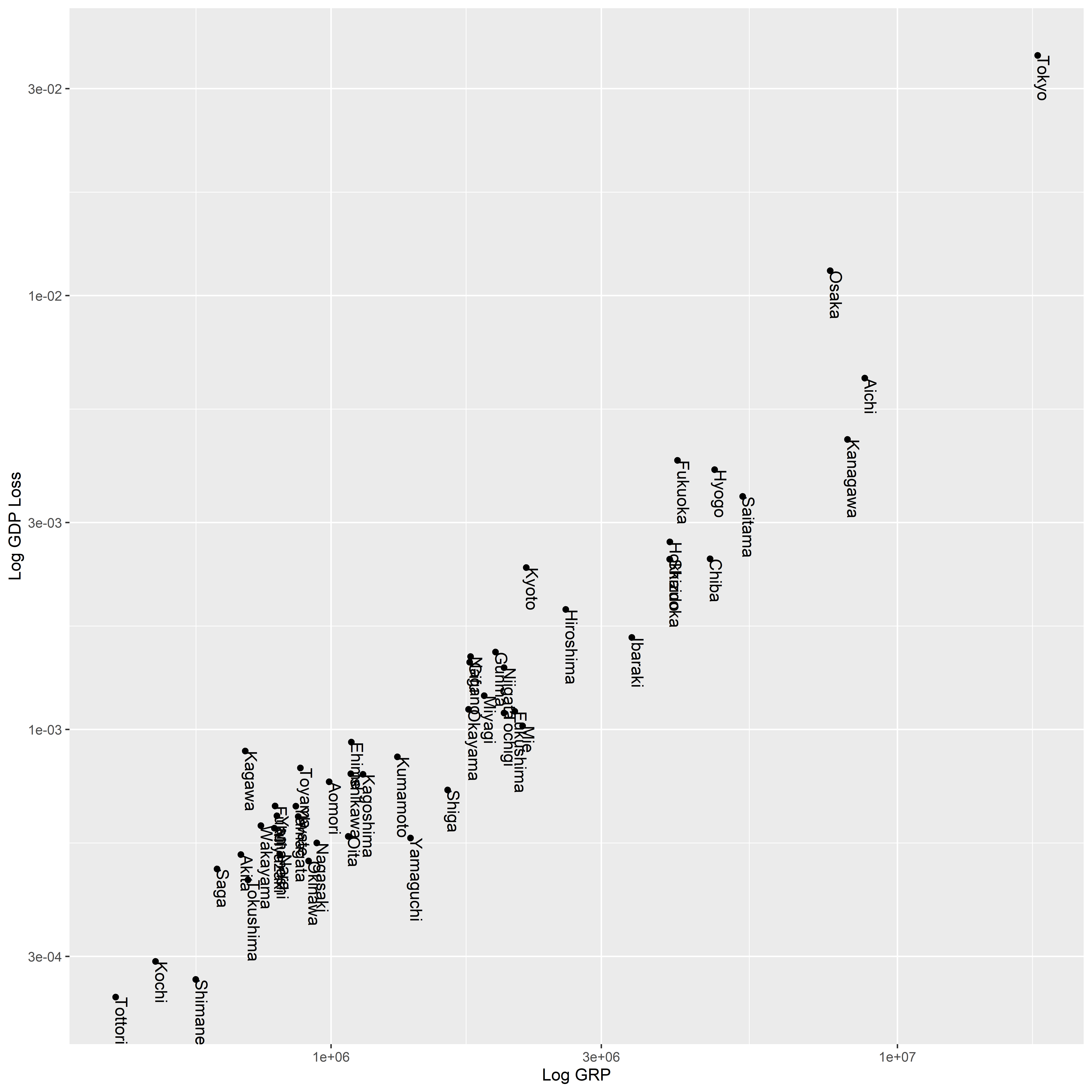}
\caption{GRP of prefectures and GDP loss. GDP loss is estimated base on one-prefecture restriction. The restriction is applied to all sectors for four weeks. The horizontal axis is the GRP of the restricted prefecture. The vertical axis is the GDP loss caused by the restriction. Some labels overlap but for small GRP, this is unavoidable, and they are not particularly important in this figure.}
\label{fig:grpvs}
\end{figure}

Figure \ref{fig:grpvs} shows the GRP of the prefecture with a restriction and the GDP loss caused by it. The restriction is for all sectors and lasts four weeks.

\begin{figure}[htbp]
    \centering
    \begin{subfigure}[t]{0.49\textwidth}
        \centering
        \includegraphics[width=\linewidth]{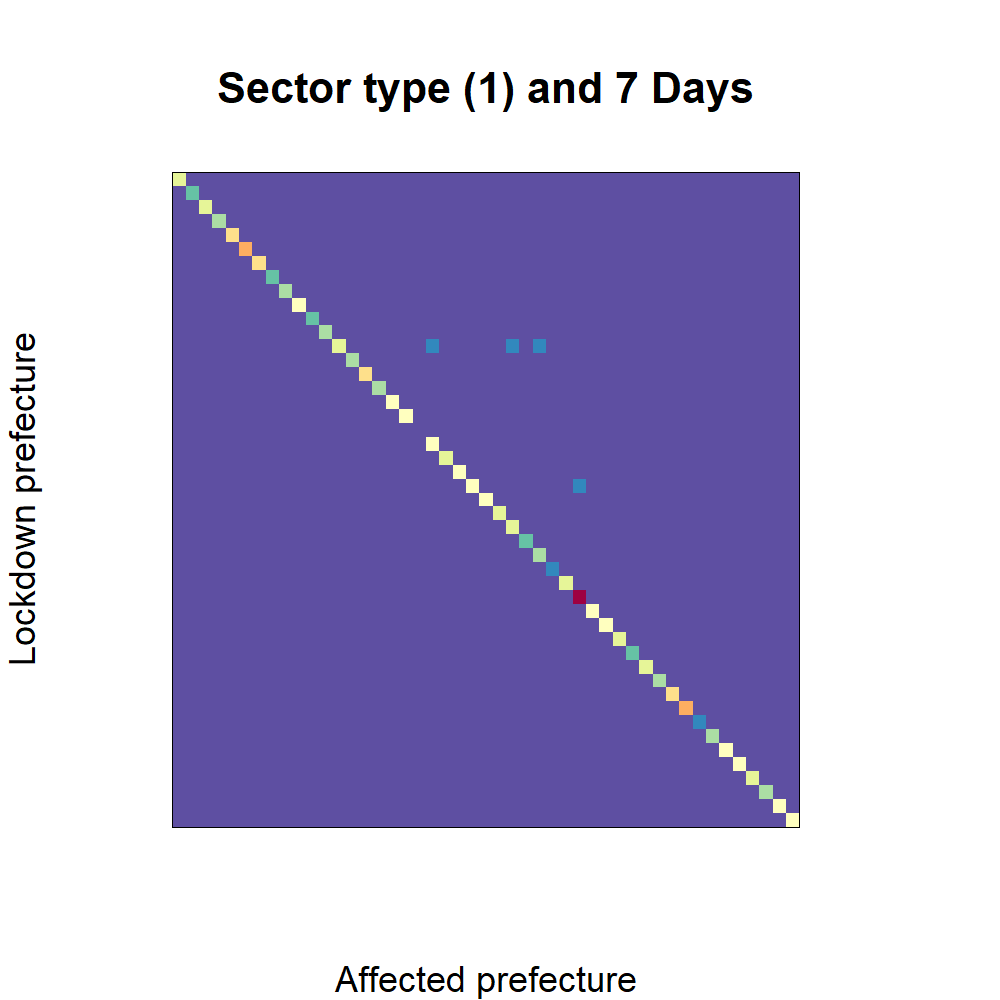} 
    \end{subfigure}
    \hfill
    \begin{subfigure}[t]{0.49\textwidth}
        \centering
        \includegraphics[width=\linewidth]{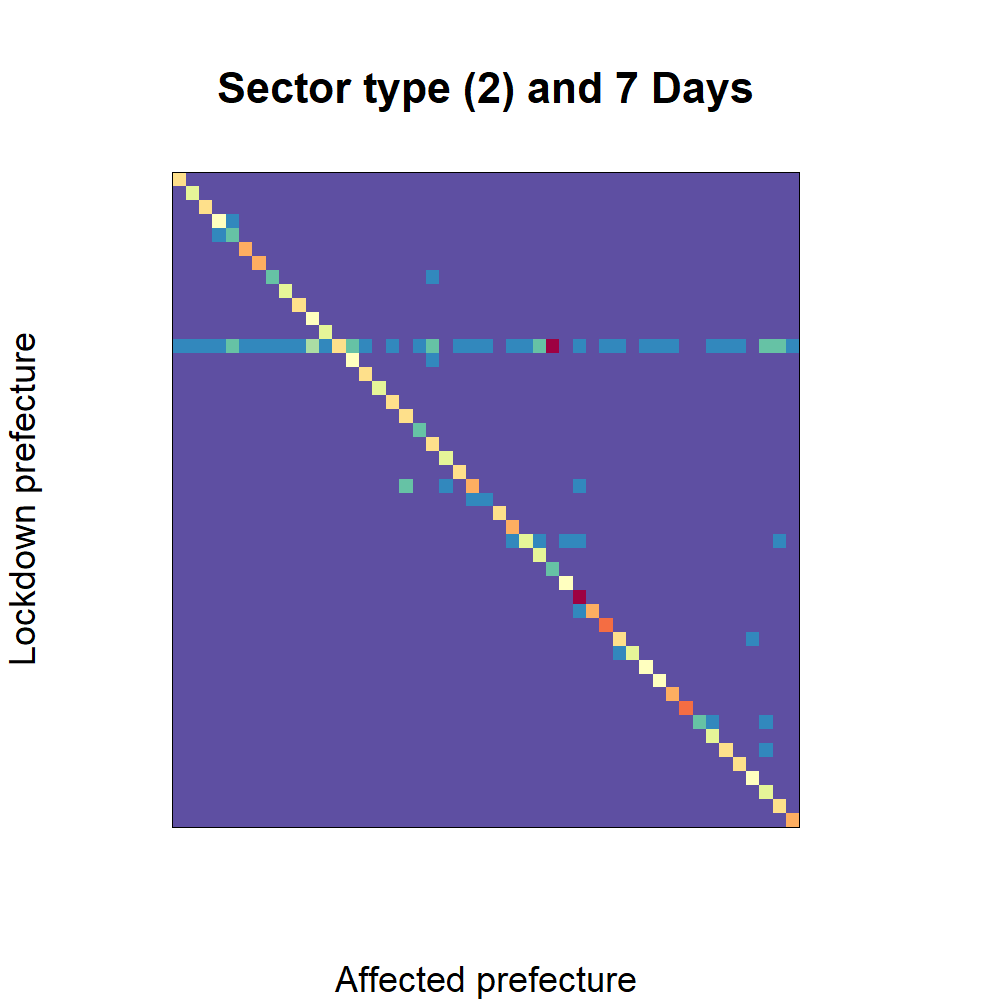} 
    \end{subfigure}
    \hfill

    \vspace{1ex}

    \begin{subfigure}[t]{0.49\textwidth}
        \centering
        \includegraphics[width=\linewidth]{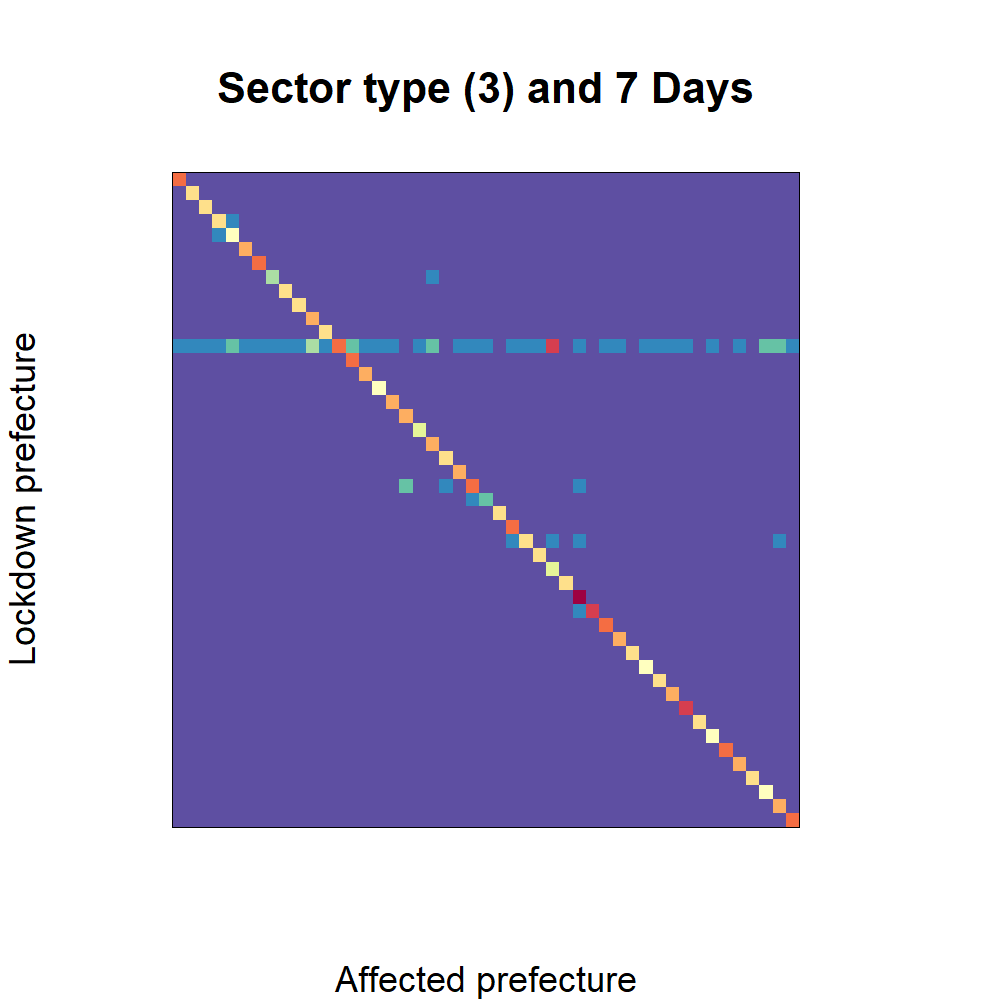} 
    \end{subfigure}
    \hfill
    \begin{subfigure}[t]{0.49\textwidth}
        \centering
        \includegraphics[width=\linewidth]{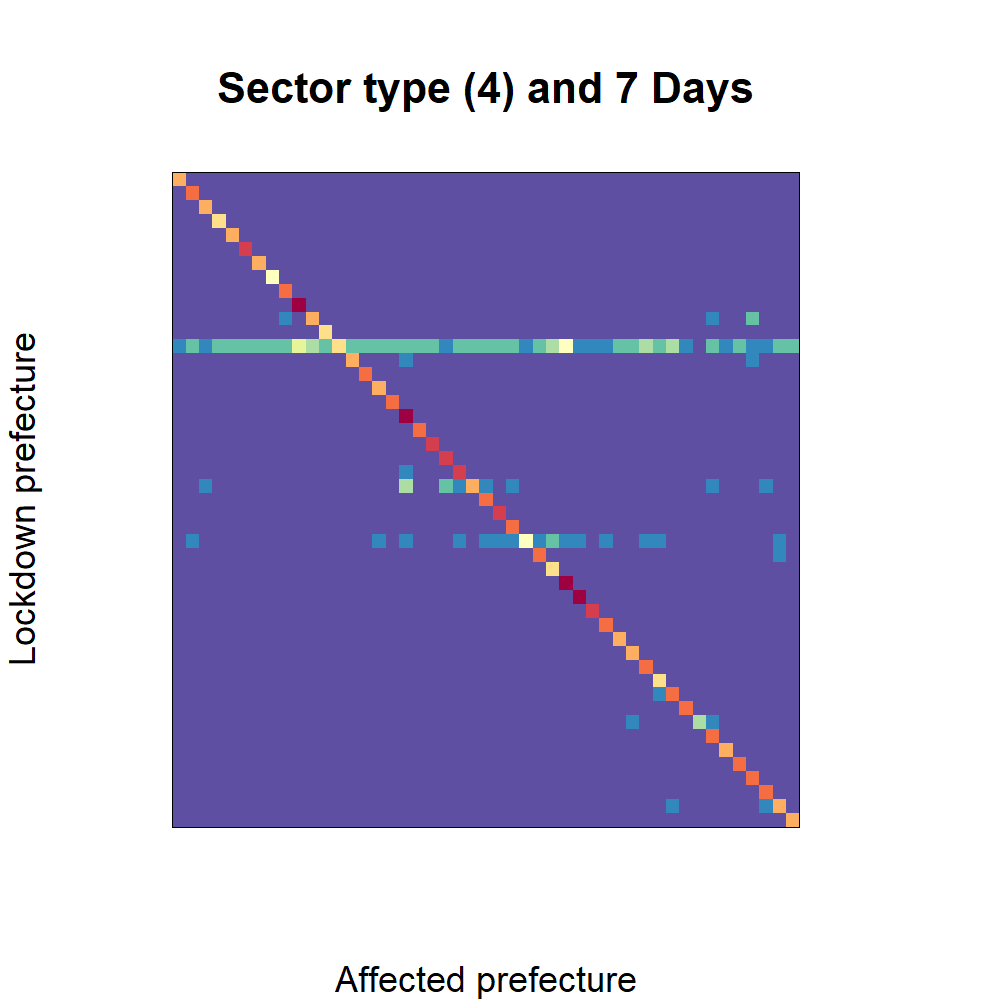} 
    \end{subfigure}

    \caption{Economic impact of regional restrictions. Each row shows a restriction in a prefecture. Each element shows an economic loss caused by the restriction in the prefecture. The magnitude of the GRP loss is shown by colors. The restriction in this figure lasts 1 weeks. The sector types are (1) the accommodation and leisure sectors, (2) the restaurant sector plus (1), (3) the retail sector plus (2), and (4) all sectors.}
    \label{fig:7heat}
\end{figure}

\begin{figure}[htbp]
    \centering
    \begin{subfigure}[t]{0.49\textwidth}
        \centering
        \includegraphics[width=\linewidth]{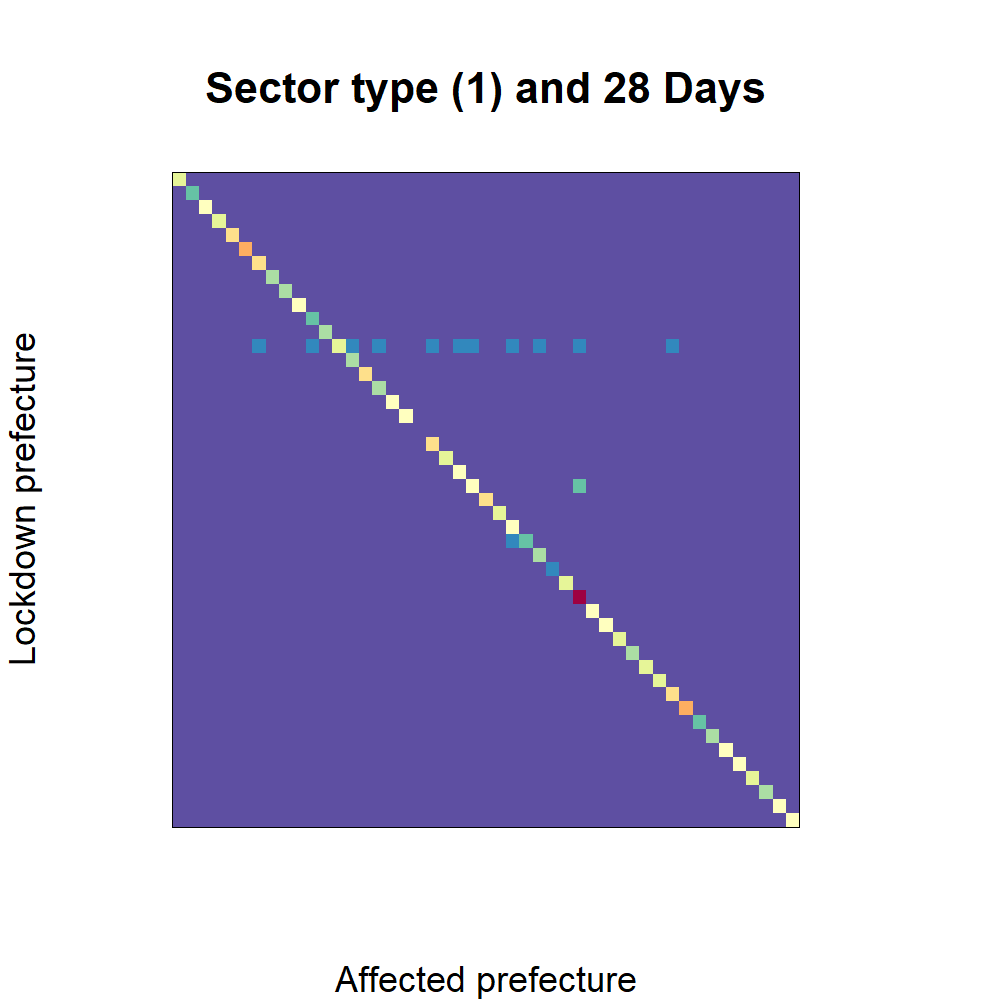} 
    \end{subfigure}
    \hfill
    \begin{subfigure}[t]{0.49\textwidth}
        \centering
        \includegraphics[width=\linewidth]{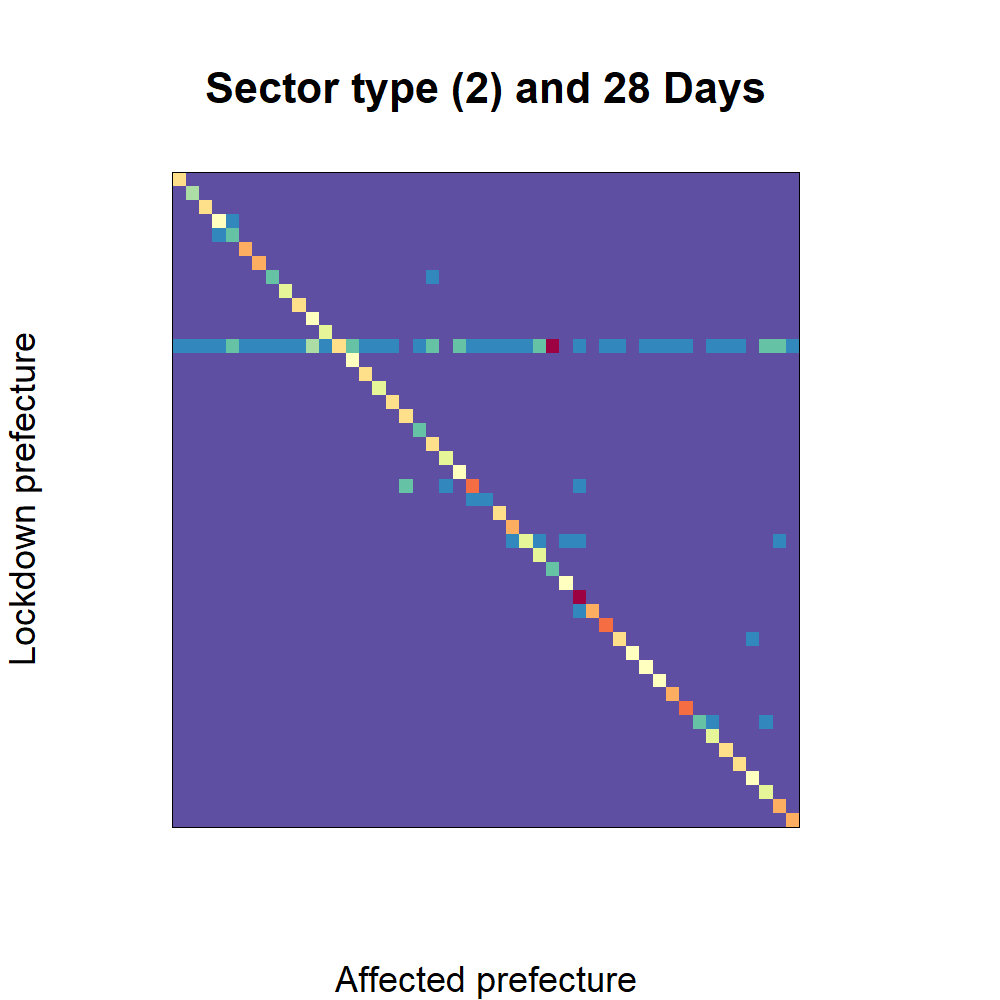} 
    \end{subfigure}
    \hfill

    \vspace{1ex}

    \begin{subfigure}[t]{0.49\textwidth}
        \centering
        \includegraphics[width=\linewidth]{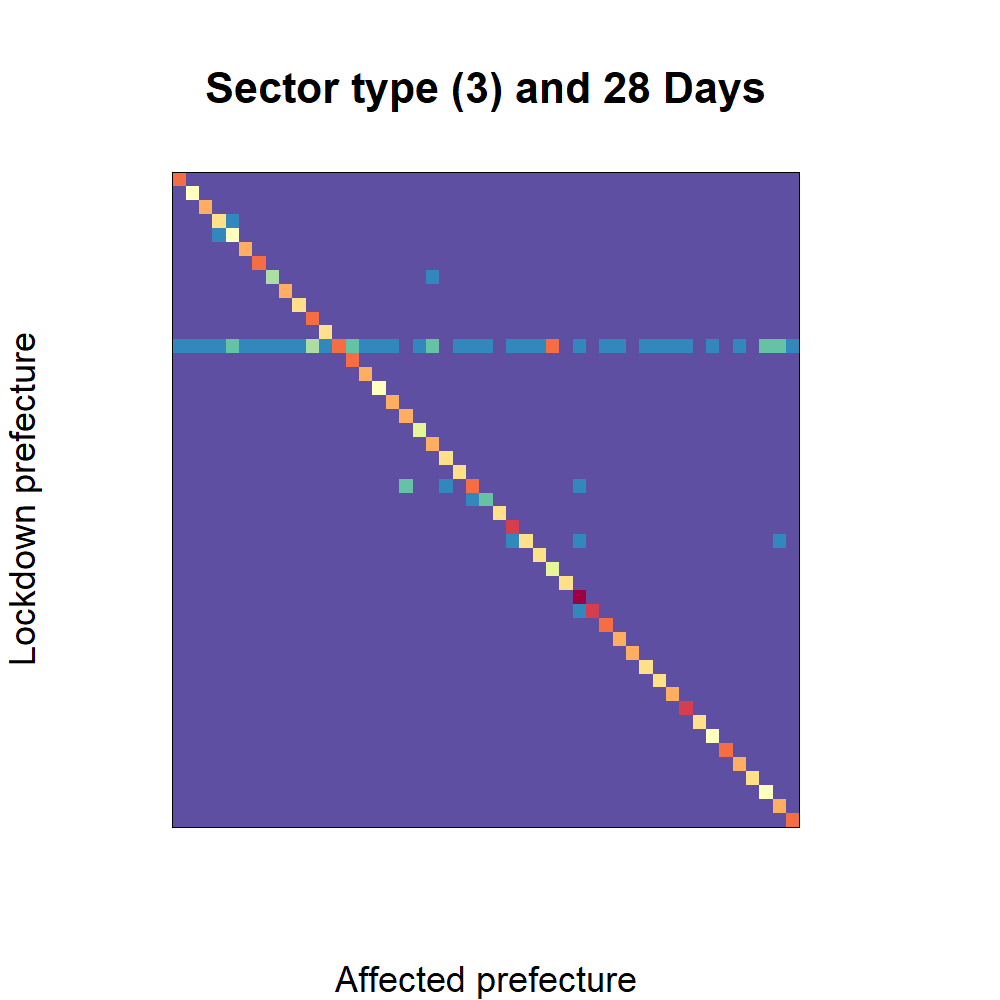} 
    \end{subfigure}
    \hfill
    \begin{subfigure}[t]{0.49\textwidth}
        \centering
        \includegraphics[width=\linewidth]{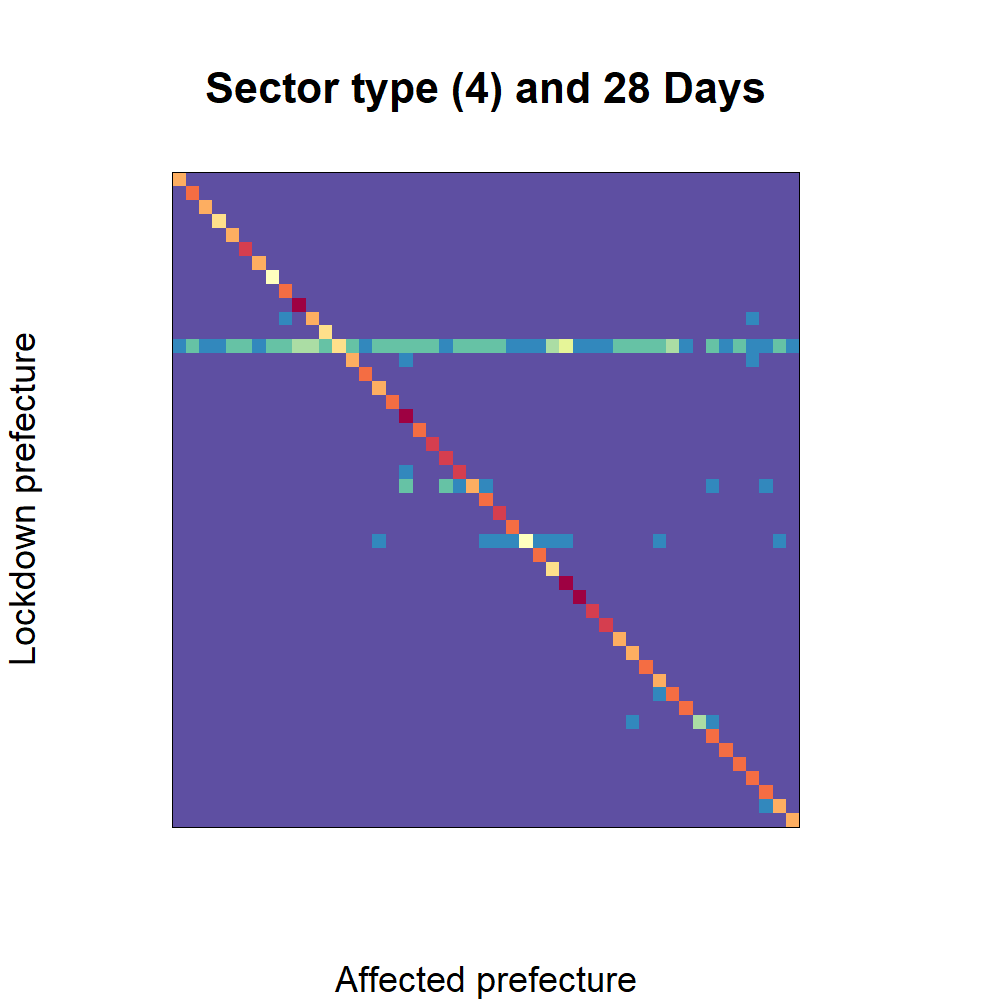} 
    \end{subfigure}

    \vspace{1ex}

    \caption{Economic impact of regional restrictions. Each row shows a restriction in a prefecture. Each element shows the economic loss caused by the restriction in the prefecture. The magnitude of the GRP loss is shown by colors. The restriction in this figure lasts four weeks. The sector types are (1) the accommodation and leisure sectors, (2) the restaurant sector plus (1), (3) the retail sector plus (2), and (4) all sectors.}
    \label{fig:28heat}
\end{figure}

Figures \ref{fig:7heat} and \ref{fig:28heat} show the effects of the one-prefecture restrictions for one and four weeks, respectively. Although we test two- and three-week durations as well, there are few differences from the results based on one- and four-week durations, and thus we omit them. The sector levels are (1) the accommodation and leisure sectors, (2) the restaurant sector plus (1), (3) the retail sector plus (2), and (4) all sectors.

\clearpage

\subsection{Effects of two-prefecture restrictions}
\label{ch:twosi}

\begin{figure}[htbp]
\centering
\includegraphics[width=\linewidth]{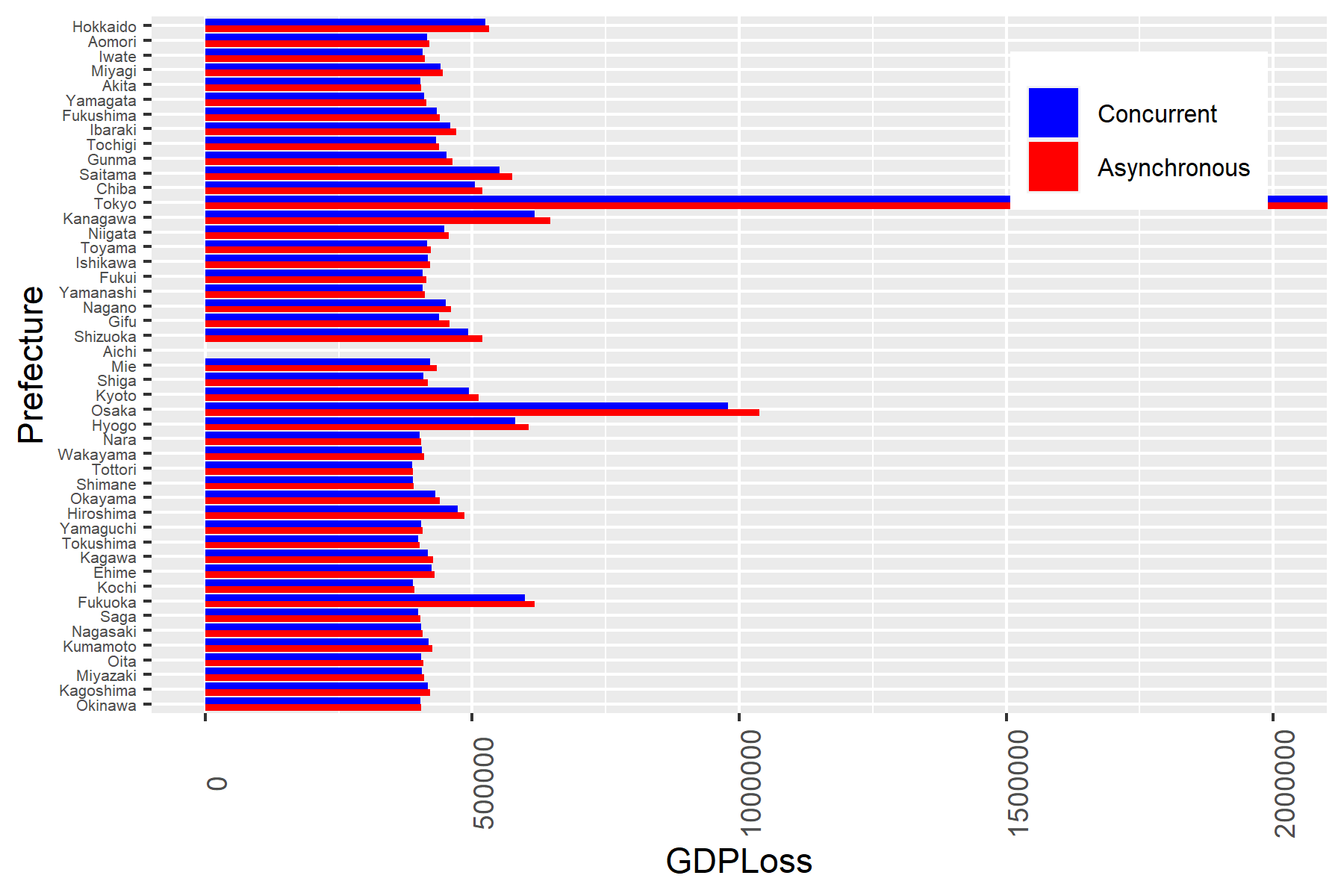} 
    \caption{Comparisons of asynchronous and concurrent simulations of restrictions in the specified prefecture and Aichi. The restrictions are imposed on the two prefectures in each simulation, with the prefectures with restrictions indicated on the vertical axis. The restrictions are imposed on all sectors for four weeks. The blue bars show the outcomes of the asynchronous simulations, that is, the GDP losses from the two independent simulations, whereas the red bars show the GDP losses from the concurrent simulations. The bars are the average of 30 Monte Carlo simulations.}
    \label{fig:aichibar}
\end{figure}

\begin{figure}[htbp]
\centering
\includegraphics[width=\linewidth]{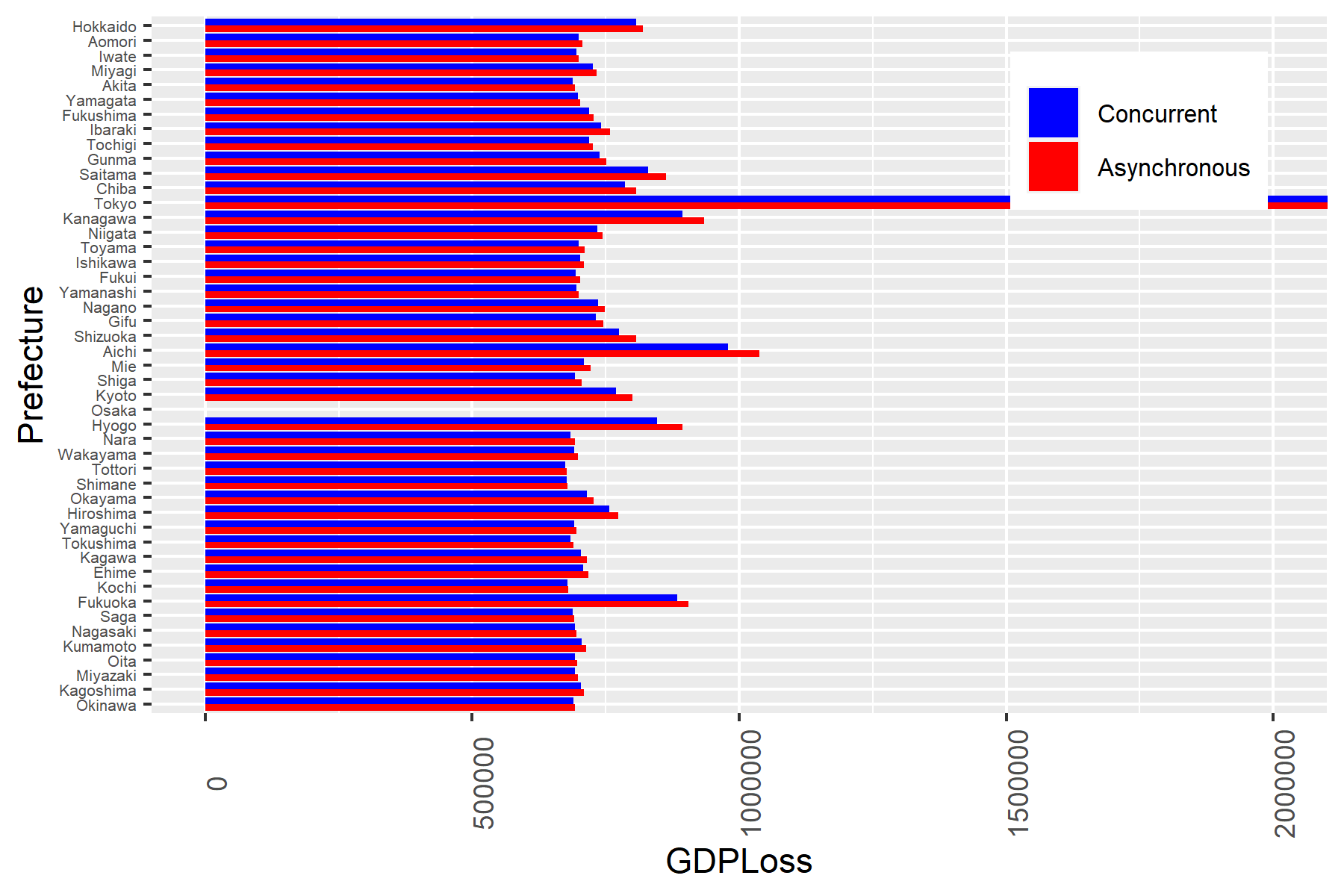} 
    \caption{Comparisons of asynchronous and concurrent simulations of restrictions in the specified prefecture and Osaka. The restrictions are imposed on the two prefectures in each simulation, with the prefectures with restrictions indicated on the vertical axis. The restrictions are imposed on all sectors for four weeks. The blue bars show the outcomes of the asynchronous simulations, that is, the GDP losses from the two independent simulations, whereas the red bars show the GDP losses from the concurrent simulations. The bars are the average of 30 Monte Carlo simulations.}
    \label{fig:osakabar}
\end{figure}

\begin{figure}[htbp]
\centering
\includegraphics[width=\linewidth]{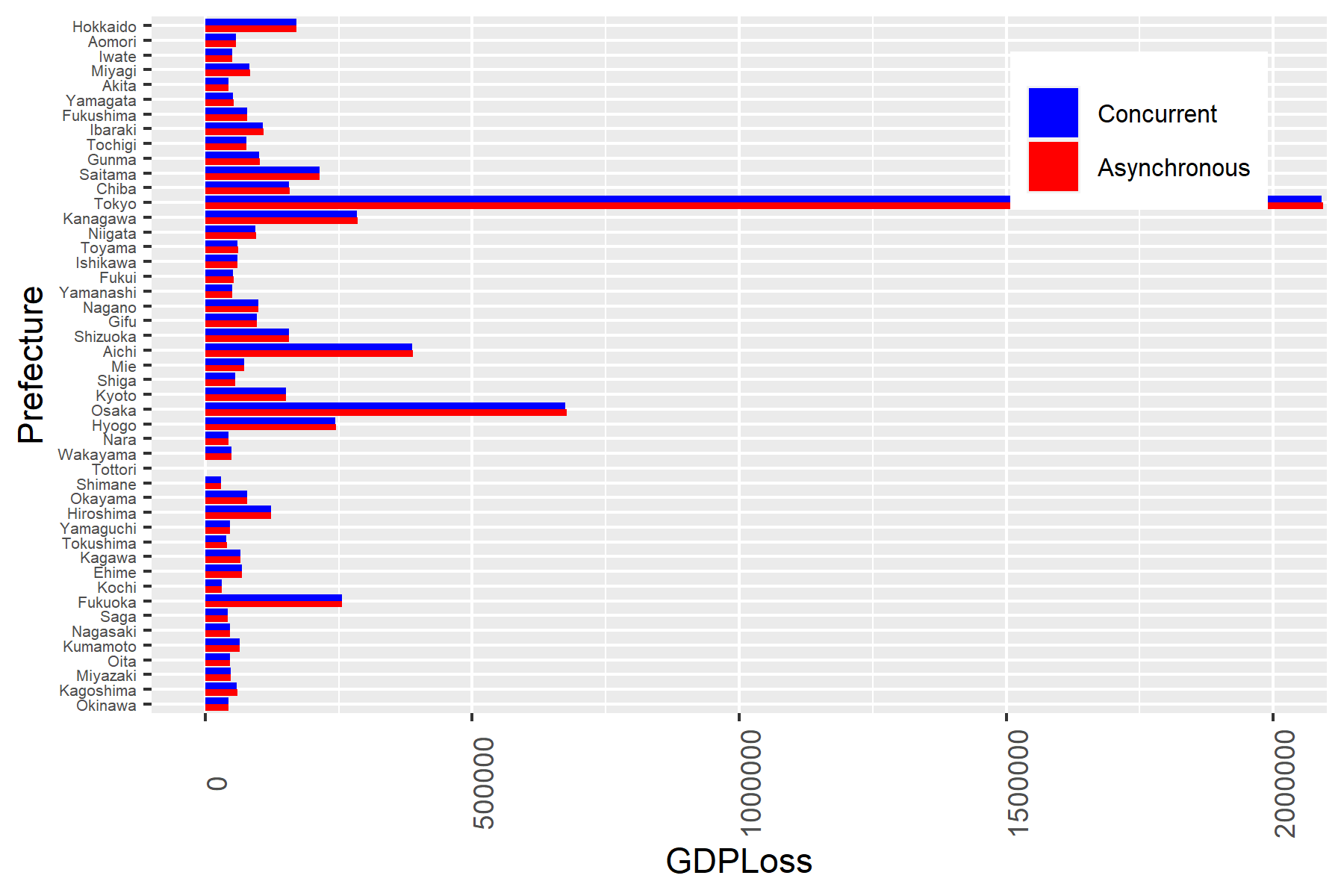} 
    \caption{Comparisons of asynchronous and concurrent simulations of restrictions in the specified prefecture and Tottori. The restrictions are imposed on the two prefectures in each simulation, with the prefectures with restrictions indicated on the vertical axis. The restrictions are imposed on all sectors for four weeks. The blue bars show the outcomes of the asynchronous simulations, that is, the GDP losses from the two independent simulations, whereas the red bars show the GDP losses from the concurrent simulations. The bars are the average of 30 Monte Carlo simulations.}
    \label{fig:tottoribar}
\end{figure}

Figures \ref{fig:aichibar}, \ref{fig:osakabar}, \ref{fig:tottoribar} show the results of the two-prefecture restrictions, in which one of the two prefectures is Aichi, Osaka, or Tottori, respectively.

\clearpage

\subsection{Effects of nationwide coordination of restrictions}

\begin{figure}[htb]
\centering
\includegraphics[width=0.8\linewidth]{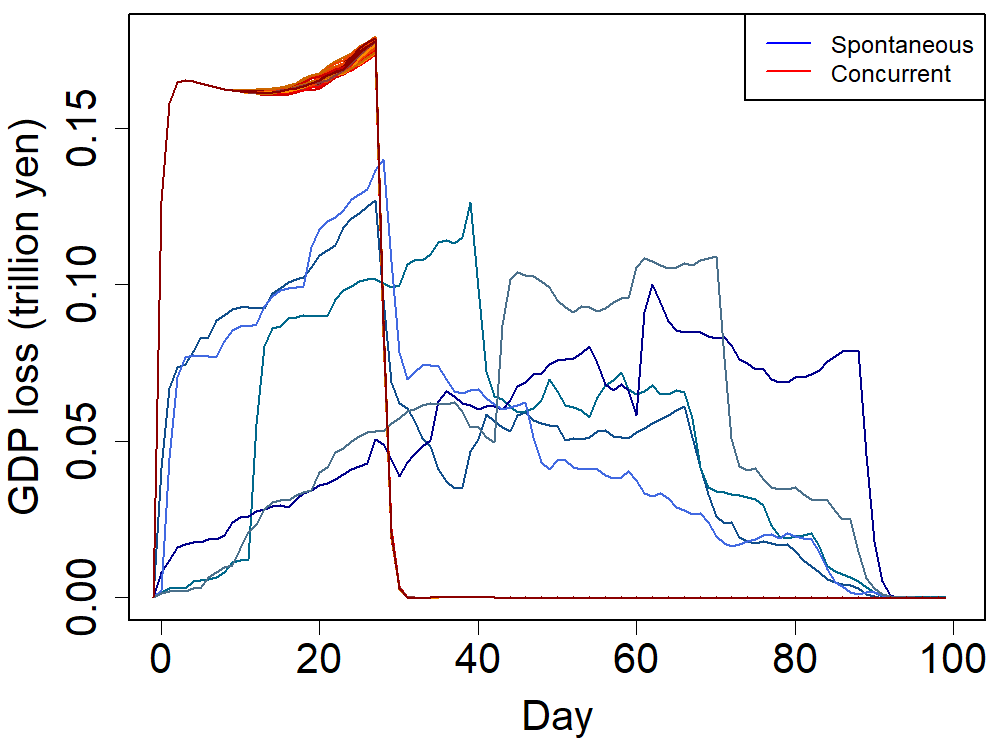}
\caption{Five random samples of the asynchronous simulations from Figure \ref{fig:congestionAll}. The vertical axis shows GDP. The horizontal axis shows days. The blue lines are the five random samples from Figure \ref{fig:congestionAll}. They show asynchronous restrictions in the prefectures with the four-week restriction timings randomly chosen from a range of three months. The red lines show four-week concurrent restrictions for all prefectures.}
\label{fig:congestionAllEx}
\end{figure}

Figure \ref{fig:congestionAllEx} shows five random asynchronous samples from Figure \ref{fig:congestionAll}.

\end{document}